\documentclass[12pt]{article}

 \usepackage{graphicx} 
\usepackage{stackengine}\setstackEOL{\!}
\usepackage{setspace} 
\usepackage{dcolumn} 
\usepackage{appendix}
\usepackage{comment}
\usepackage{multirow,tabularx}
\usepackage{threeparttable}
\usepackage{float}
\usepackage{lipsum}
\usepackage{csquotes}
\usepackage[normalem]{ulem}
\usepackage[section]{placeins}
\usepackage{bibunits}
\usepackage{natbib} 
\usepackage{multicol}
\usepackage[normalem]{ulem}
\usepackage{amsmath} 
\usepackage{venndiagram}
\usepackage{amscd} 
\usepackage{amsmath} 
\usepackage{amsthm}
\usepackage{amscd} 
\usepackage{mathpazo} 
\usepackage{sgame} 
\usepackage{comment} 
\usepackage{amsmath} 
\usepackage{lscape}
\usepackage[hyphens]{url} 
\usepackage{amscd} 
\usepackage{sgame} 
\usepackage{caption}
\usepackage{pstricks} 
\usepackage[hyphens]{url} 
\usepackage{graphicx} 
\usepackage{mathtools}
\usepackage{ushort}
\usepackage{censor}
\newtheorem{theorem}{Theorem}
\newtheorem{proposition}{Proposition}
\newtheorem{corollary}{Corollary}

  \defaultbibliography{bibliography.bib}
  \defaultbibliographystyle{aer}

\usepackage{relsize,etoolbox}
\AtBeginEnvironment{quote}{\small\it}

\def\citepos#1{\citeauthor{#1}'s \citeyear{#1}} 

\usepackage{hyperref} \hypersetup{
    colorlinks = true,
    linkcolor = red,
    anchorcolor = red,
    citecolor = blue,
    filecolor = red,
    urlcolor = blue
}
  \defaultbibliography{Writeup/political-influence-concentration}
  \defaultbibliographystyle{aer}

\begin{document}
\begin{bibunit}
\singlespacing

\title{\huge Political Power and Market Power\thanks{We thank Laura Alexander, Peter Scott Campbell, Paola Conconi, Caroline Flammer, Dana Foarta, Jens Frankenreiter, Simon Freyaldenhoven, Jorge Guzman, Aseem Kaul, Daniel Keum, In Song Kim, Ameet Morjaria, Fiona Scott Morton, Patryk Perkowski, Noemie Pinardon-Touati, Daniela Scur, Haram Seo, Reed Showalter, Cailin Slattery, Steve Tadelis, Glen Weyl, Brian Wu, Tim Wu, Mark Xu, and seminar participants in Barcelona, Brussels, Chicago, Columbia, HEC, Kellogg (Northwestern), K\"{o}ln, Norwegian School of Economics (Bergen), SSES Fribourg, and Stanford for helpful suggestions. We also thank conference participants at NBER Summer Institute (Political Economy), NBER Megafirms, the Society for Institutional and Organizational Economics (SIOE), Strategy and the Business Environment (SBE), USMA (West Point) and the Wharton Corporate Strategy and Innovation Conference. We also thank Bettina Hammer, Jett Pettus, Szymon Sacher and Natalie Yang for research assistance, and Fabrizio Dell'Acqua for contributions to an earlier version of this project.}}

\author{{\large{}Bo Cowgill}\\
\emph{\small{}Columbia University}\\
\and{\large{}Andrea Prat}\\
\emph{\small{} Columbia University}\\
\and{\large{} Tommaso Valletti}\\
 {\small{}
}\emph{\small{} Imperial College London}{\small{}}\\
{\small{}{}} }

\date{March 2023\phantom{s}(First Version: June 2021) \\
\vspace{10mm}
\href{https://papers.ssrn.com/sol3/papers.cfm?abstract_id=4390776}{\textbf{Click here for the most recent version.}}}

\maketitle

\begin{abstract}
We study the link between political influence and industrial concentration. We present a joint model of political influence and market competition: an oligopoly  lobbies the government over regulation, and competes in the product market shaped by this influence. We show broad conditions for mergers to increase lobbying, both on the intensive margin and the extensive margin. We combine data on mergers with data on lobbying expenditures and campaign contributions in the US from 1999 to 2017. We document a positive association between mergers and lobbying, both by individual firms and by industry trade associations. Mergers are also associated with extensive margin changes such as the formation of in-house lobbying teams and corporate PACs. We find some evidence for a positive association between mergers and higher campaign contributions. \\ 

\end{abstract}


\clearpage
\setstretch{1.5}

\section{Introduction}

Lobbying and campaign finance are an essential element of modern representative democracy \citep{grossmanhelpmanbook,ansolabeheredefiguesnyder,cageprice}. On the positive side, they can help elected officials to gather information needed to make legislative and regulatory choices, and can help voters become informed about candidates on the ballot. However, they also both raise legitimacy and fairness concerns, as individuals and organizations with greater wealth can spend more and exercise greater influence over the political process. 

In this paper, we study the link between lobbying and concentration in industries. This link is important for two reasons. First, businesses represent the largest source of lobbying spend. According to data from OpenSecrets, businesses accounted for 87 percent of total lobbying spending in the US in 2019 and 36 percent of contributions from Political Action Committees (PACs) in the 2017/18 political cycle (where labor and ideological contributions also play a big share). 

Second, in recent years there has been rising concern that industrial concentration not only directly affects consumers  through market power (potentially raising prices and reducing quantities), but indirectly affects consumers through politics \citep{zingales2017towards, wu2018curse}. Concern over the political influence of concentrated industries has appeared throughout the history of antitrust \citep[e.g.][]{brandeis1914other, pitofsky1978political, khan2017ideological}.\footnote{One example of this is Thomas Jefferson sought to add ``freedom from monopolies'' to the Bill of Rights in the U.S. Constitution \citep{jefferson1789thomas}.} 
Incumbent firms could lobby politicians to erect barriers to entry and protect their market power. This is another form of consumer harm, but the channel through which it flows is regulation. If lobbying exhibits economies of scale, an increase in market concentration should lead to an increase in lobbying activity. If this hypothesis is correct, market power begets political power. 

We begin with a brief theoretical section that shows how even a very simple model provides ambiguous predictions as to whether mergers should increase or decrease influence activities. We consider an oligopoly in which firms' profits may be affected by government regulation. Firms engage in lobbying activity according to the menu auction model developed by \cite{grossman1994protection}. 

We characterize the joint equilibrium in the product market and in the ``political market'', as well as the effect of a merger between two competitors on such equilibrium. A merger can lead to an increase in lobbying activity -- for instance when regulation is a common good for the incumbent firms that operate in the industry under consideration -- or to a decrease in lobbying activity -- for instance when regulation is a pure private good in that industry.

The core of the paper studies whether in the US mergers are associated to an increase or a decrease in influence activities. We use data from SEC-registered companies in 1999-2017 (using Compustat). We match these companies with information on federal lobbying data and on campaign contributions in the US. Finally, we have detailed information about M\&A transactions over the same period. We first document how political influence spending occurs within and across industries, showing a positive relationship between relative size of a firm and its spending on lobbying and campaign contributions. 

Then, we focus on how political influence spending varies before and after a merger. We pursue two empirical approaches, both based on the timing of mergers. In the first, we use a panel event study design \citep{gentzkow2011effect, de2020two, freyaldenhoven2021visualization, goodman2021difference,  athey2022design}. Qualitatively, identification in this approach relies on the idea that mergers are endogenous, but depend on fixed (or slow-moving) variables whose trends we control for. The identification assumption is that the timing of the mergers, after conditioning on other factors, comes from idiosyncratic shocks that are unrelated to the returns of political spending. 

Our second identification strategy is a differential exposure design \citep{borusyak2020non, goldsmith2020bartik, breuer2021bartik} that uses logic similar to the \cite{bartik1991benefits} instrumental variable design. Like other Bartik-like designs, ours uses a combination of time-varying shocks and initial characteristics of companies. For shocks, we use the well-documented pattern of mergers arriving in waves \citep{nelson1959merger, gort1969economic, weston1990mergers}. These waves span multiple sectors and have several proposed causes ranging from macroeconomic shocks to technology shocks. We utilize economy wide pro-merger shocks at different times to construct a time-varying instrument similar to the \cite{bartik1991benefits} instrument.

In both designs, our empirics suggest that mergers are positively associated with an increase in firms' spending on political influence activities. The average merger is associated with a \$74K to \$106K increase in the amount spent on lobbying per period (half year) after the merger, or approximately 22\% of average per-period spend of merging firms. The average merger is also associated with an approximately \$4K to \$10K increase in campaign contributions per period, but this association is not statistically significant in all specifications. 

In most specifications, the association of mergers with influence activities appears to be significantly stronger if the merging companies are larger and if the merging companies belong to the same industry. 

We also consider a possible mis-specification problem. Merging firms may ramp up their influence activities \emph{before} the merger, possibly to increase the chance of the transaction being approved by regulatory authorities. However, we find no evidence in the data for such an anticipation effect. This null result may be a reflection of the fact that most mergers during our sample period were not scrutinized by US antitrust authorities \citep{wu2018curse}.

\subsection{Related Research}

Our paper aims to contribute to two main lines of research in political economy. 

\paragraph{Theories of Political Influence.} First, we contribute a novel political economy model of the relationship between political outcomes and marketplace dynamics. This topic has been the focus of many researchers outside of economics \citep[e.g.][ and others]{brandeis1914other, pitofsky1978political, khan2017ideological, wu2018curse}. Within economics, models by \cite{tullock1967welfare, stigler1971theory, hillman1982declining} and \cite{mcchesney1987rent} formalize early ideas of regulation as a function of industry influence. We follow that literature in using \citepos{grossman1994protection} model as the basis for our theoretical approach. A recent model by \cite{bombardini2012competition} studies why highly competitive industries could nonetheless cooperate on lobbying. \cite{huneeus2018effects} studies the relationship between firm size and lobbying, and the resulting misallocation of firm resources.

\cite{callanderfoartasugaya} develop an integrated dynamic model of competition, innovation, and policy-making. They show the existence of a feedback loop between market power and political power. In equilibrium, the policy-maker ``manages competition'' to protect the incumbent, resulting in less competition and innovation.

Our theoretical approach has two distinguishing features. First, we allow a firm's willingness to lobby to arise endogenously in response to the business and political environment, including in response to mergers. Second, we allow lobbying not only to affect policy, but also to influence prices and quantities through regulation. These are often modeled separately, while we combine them into a single integrated model featuring two blocks (a model of industrial organization model of competition under regulation, as well as a political economy model of lobbying for the regulation).\footnote{A notable exception is \cite{bombardini2012competition}, which studies the formation of industry associations as a function of how competitive product markets are for an industry (assuming Bertrand competition of differentiated goods).} Our blending of these models creates the potential for feedback loops between product markets and politics.  

We allow for multiple types of industry regulation. Much of the prior literature both in theory and empirics is motivated by trade, where domestic firms are typically united in their preference for protection. This creates free rider problems which are present in our model in line with earlier papers \citep{olson1965logic, grossman1994protection}.\footnote{Freeriding and ``public good'' aspects of lobbying appear outside of economics as well e.g. \cite{baumgartner1998basic, hart2004business, barber2014lobbying}.} However, we also allow regulation to divide competitors by helping some  at the expense of others. This is  particularly important for the political economy of antitrust. This would apply, e.g., when a market leader lobbies for regulations to protect its position, while a challenger opposes the regulations (and/or prefers others). 
Should the incumbent merge with the challenger, this form of rivalrous lobbying would disappear. Although our data come from a developed economy within a democratic state (the U.S., 1999-2017), our model is not specific to a type of government. Similar business/government dynamics could appear under other institutional arrangements. State capture by business interests is a theme appearing in development economics \citep{canen2022political}.

\paragraph{Empirical Studies of Special Interest Politics.} Our paper also contributes to the empirical literature. Our analysis is related to a small but growing set of studies linking industry-level variables with lobbying activities.\footnote{For a survey of the empirical literature on lobbying see \cite{bombardini2020empirical}.} The pioneering work in the area is \cite{goldberg1999protection}, which tests and estimates \citepos{grossman1994protection} model with industry-level US data on lobbying and tariffs. A set of recent related papers study in particular how lobbying tries to influence trade agreements \citep[e.g.,][]{bombardini2012competition, blanga2021lobbying}. Many of the prior studies conduct cross-sectional comparisons between firms or industries; a key feature of our empirical approach is the use of within-industry and even within-firm changes in merger status over time. 

\cite{bombfrontier2021} study lobbying in the US as a consequence of imports from China, showing differential responses between firms on the technological frontier and laggards. \cite{bertrand2020investing} study the effect of the identity of a firm's shareholders on that firm's campaign contribution patterns. The probability that a firm's PAC donates to a politician supported by an investor's PAC doubles after the investor acquires a large stake. Like ours, this study uses changes within the same firm over time (in their case, changes to ownership). 

A series of recent empirical papers documents firm mark-ups, higher aggregate industry concentration, a decline in the labor share of output, larger firm and income inequality, and a reduction in business dynamism \citep{philippon2019great,de2020rise,dube2020monopsony}. \cite{reed2021} and \cite{McCarty2021} show these trends were concurrent with increases in lobbying and industry concentration. Our paper aims to connect these trends more directly, both using a theoretical model of lobbying and concentration, as well as through causal empirical evidence about the linkage between concentration and political influence. Our empirics are particularly related to the political economy of antitrust. \cite{mehta2020politics} and \cite{fidrmuc2018antitrust} measure political interference in the antitrust review process from members of Congress and corporations. Instead, we focus on the impact that merger policy can have on lobbying for regulation more generally. 

\paragraph*{Mergers.} Finally, we contribute some innovations to the study of mergers and acquisitions. From a firm's perspective, our results speak to a novel type of merger benefit: ``non-market synergies'' such as coordinated activity in government affairs \citep{baron1995integrated, feldman2021synergy}. We show how this form of merger benefit arises from externalities in the (uncoordinated) non-market choices of competing firms, and how the benefit of coordination can either increase or reduce overall lobbying. Our results show an example of a non-market strategy (lobbying to erect regulatory barriers to entry) complementing a marketplace strategy (merging and setting prices and quantities in product markets). 

We also contribute methodological innovations about mergers. Our research questions require us to examine a bundle of firms as a single unit, and measure the bundle's aggregate characteristics over time (such as before/after mergers). To our knowledge this is a distinctive approach in the literature on mergers. 
For identification, one of our strategies uses a differential exposure design \citep{borusyak2020non, goldsmith2020bartik, breuer2021bartik}, using logic similar to the \cite{bartik1991benefits} instrument. Similar Bartik-like designs have been deployed to study local labor effects of Chinese trade \citep{david2013china}, native/immigrant substitution \citep{card2009immigration} and credit shocks during the Great Recession \citep{greenstone2020credit}. We propose and execute an adaptation of this strategy to examine merging firms.  

While very much related in spirit to many of the papers above, to our knowledge, ours is the first paper that attempts to link, both theoretically and empirically, the industrial concentration induced by mergers with lobbying activities and PAC spend. The next section presents our theory, and Sections \ref{sec:empoverview} and \ref{sec:data} provide an overview of our empirical approach and data. Sections \ref{sec:panelevent} through \ref{sec:altexplan} present our empirical strategies and results, and Section \ref{sec:conclusion} concludes. 





\section{Theory}

In this section we present a simple model of lobbying and competition. This model is composed of two building blocks: an industrial organization model of oligopoly with regulation, and a political economy model of lobbying for regulation. Our aim is to analyze how the equilibrium in the lobbying game is affected by a merger in the industry. To show the forces at play, we discuss the simplest possible setting: an initial duopoly, to be assessed against a merger to monopoly. Proofs for propositions are in Appendix \ref{app:proofs}. 

\subsection{Competition}

We begin with the industrial organization block, which we take to be a standard quantity competition model augmented with regulatory variables.\footnote{We use a Cournot setting as it is the one with the simplest analytical expressions one can obtain. A merger to monopoly is profitable and does not suffer from the merger paradox of Cournot games with more than two firms.} Consider an industry with initially 2 firms. Each firm can set its own quantity $q_{i}$, as well as lobby for some regulations. Each firm can lobby over two dimensions: a common component that is favorable to the whole industry, that we denote as $R$, and a private component that is favorable only to the specific firm, and that we denote as $F_{i}.$ The resulting demand for firm $i$ is assumed to be linear and equal to%
\[
P_{i}=A+aR+bF_{i}-Q,
\]
where $Q=q_{1}+q_{2}$ and $A$ is a proxy for industry size.

This is a standard Cournot model that has been augmented with regulatory variables. The term $aR$ represents the common effect of regulation on demand. We can think of $R\in\Re$ as government policy that is favorable to the industry. An increase in $R$ increases demand for the incumbent firms, everything else equal. In particular, $R$ can be thought of as the result of an additional cost $\tau$ imposed on a competing product that \textit{could} be sold in the industry. This applies, e.g., to at least two well-studied form of regulations. First, the alternative product could come from the international competition and the cost $\tau$ is an import tax, as studied in the tariff lobbying of \cite{grossman1994protection}. Second, the alternative product could be a different set of domestic producers and $\tau$ would be an explicit or implicit barrier to entry. By lobbying over $R,$ the incumbent duopolists can fend off entry from these alternative competitors by making $\tau$ sufficiently high. The parameter $a$ simply captures the effectiveness of lobbying. 

Similarly, the term $F_{i}$ $\in\Re$ represents the effect of lobbying over a dimension that favors directly only firm $i$, but not its rival. In fact, the rival firm will be indirectly disadvantaged by this type of lobbying, as firm $i$ will want to expand its output the higher $F_{i}$ is, which reduces firm $j$'s profit. The parameter $b$ again represents the effectiveness of lobbying. A simple example of this type of regulation is a firm-targeted per-unit subsidy or tax rebate. It can also be interpreted as preferential treatment in public procurement. For instance, suppose that everything else equal, the government is willing to pay a premium for procuring Firm 1's product but not for Firm 2's product. We can represent that as $F_1>0$ and $F_2=0$.

The industrial organization part of the model is completed by a linear cost function, that we normalize to zero. Firm $i$ maximizes with respect to own quantity its profit function%
\[
\pi_{i}=p_{i}q_{i}.
\]
Each firm, in addition, makes a transfer $t_{i}$ to the regulator when
lobbying, to be discussed next.

\subsection{Lobbying}

The lobbying block follows \citepos{grossman1994protection} canonical lobbying model, which in turn is based on the menu auctions studied by \cite{bernheim1986common}.\footnote{For a recent model of dynamic lobbying with market power see  \cite{callanderfoartasugaya}.}  Suppose we have $n=2$ lobbies with profit $\pi_{i}\left(
\mathbf{P}\right)  $ where $\mathbf{P}$ is a policy vector, that in our case corresponds to $\mathbf{P}=\{R,F_{1},F_{2}\}$ and $\pi_{i}$ is the profit described above. The policy maker maximizes%
\[
\sum_{i}t_{i}+w\left(  \mathbf{P}\right),
\]
where $t_{i}$ is the contribution the regulator receives from lobby $i$ and $w$ is a direct utility function, that expresses the direct benefit or cost the regulator receives from policy. We can borrow from \cite{bernheim1986common} the following useful result.

\begin{theorem}
[Bernheim-Whinston]In any coalition-proof equilibrium of this lobbying game,

(i) The policy-maker selects%
\[
\mathbf{P}^{\ast}\in\arg\max_{\mathbf{P}}\sum_{i}\pi_{i}\left(  \mathbf{P}%
\right)  +w\left(  \mathbf{P}\right)
\]

(ii) To determine the equilibrium transfers $\hat{t}_{i}$, let%
\begin{align*}
g_{i}\left(  \mathbf{P}\right)   &  =\pi_{i}\left(  \mathbf{P}\right)
-\hat{t}_{i}\\
\mathbf{P}_{-I}^{\ast}  &  \in\arg\max_{\mathbf{P}}\sum_{j\notin I}\pi
_{i}\left(  \mathbf{P}\right)  +w\left(  \mathbf{P}\right)
\end{align*}
In equilibrium, the vector $\left(  g_{i}\left(  \mathbf{P}\right)  \right)
_{i} $ lies on the upper contour of the set defined by%
\begin{equation}
\text{ for every }I\subset\mathcal{I}\text{, }\sum_{i\in I}g_{i}\left(
\mathbf{P}^{\ast}\right)  \leq\sum_{j}\pi_{j}\left(  \mathbf{P}^{\ast}\right)
+w\left(  \mathbf{P}^{\ast}\right)  -\left(  \sum_{j\notin I}\pi_{j}\left(
\mathbf{P}_{-I}^{\ast}\right)  +w\left(  \mathbf{P}_{-I}^{\ast}\right)
\right)  . \label{e:inequalities}%
\end{equation}

\end{theorem}

If we subtract $\sum_{i\in I}\pi_{j}\left(  \mathbf{P}^{\ast}\right)  $ from
both sides of (\ref{e:inequalities}) and reverse the signs, we get%
\[
\text{for every }I\subset\mathcal{I}\text{, \ \ }\sum_{i\in I}\hat{t}_{i}%
\geq\left(  \sum_{j\notin I}\pi_{j}\left(  \mathbf{P}_{-I}^{\ast}\right)
+w\left(  \mathbf{P}_{-I}^{\ast}\right)  \right)  -\left(  \sum_{j\notin I}%
\pi_{j}\left(  \mathbf{P}^{\ast}\right)  +w\left(  \mathbf{P}^{\ast}\right)
\right)  ,
\]
which constitutes a system of inequalities that puts an lower bound on the value of the vector of transfers $\hat{t}$.

In other words, the regulator chooses the policy vector that maximizes a weighted average of direct utility and profits (we have assumed equal weights). Additionally, and importantly for our application as the lobbying expenditure
is what we observe in the data, the transfers of each firm are constrained by what the regulator could do in the alternative coalitions that do not include such firms.

We can directly specialize this general result to our case.

\begin{corollary}
With $n=2$, in any coalition-proof equilibrium%
\[
\mathbf{P}^{\ast}\in\arg\max_{\mathbf{P}}\sum_{i=1}^{2}\pi_{i}\left(
\mathbf{P}\right)  +w\left(  \mathbf{P}\right)
\]
and the transfers must satisfy%
\begin{align*}
\hat{t}_{1}  &  \geq\pi_{2}\left(  \mathbf{P}_{\left\{  2\right\}  }^{\ast
}\right)  +w\left(  \mathbf{P}_{\left\{  2\right\}  }^{\ast}\right)  -\left(
\pi_{2}\left(  \mathbf{P}^{\ast}\right)  +w\left(  \mathbf{P}^{\ast}\right)
\right) \\
\hat{t}_{2}  &  \geq\pi_{1}\left(  \mathbf{P}_{\left\{  1\right\}  }^{\ast
}\right)  +w\left(  \mathbf{P}_{\left\{  1\right\}  }^{\ast}\right)  -\left(
\pi_{1}\left(  \mathbf{P}^{\ast}\right)  +w\left(  \mathbf{P}^{\ast}\right)
\right) \\
\hat{t}_{1}+\hat{t}_{2}  &  \geq\max_{\mathbf{P}}w\left(  \mathbf{P}\right)
-w\left(  \mathbf{P}^{\ast}\right).
\end{align*}

\end{corollary}

To provide a closed-form solution, we finally posit that the direct utility function is given by
\[
w(\mathbf{P)=}-w_{1}\frac{R^{2}}{2}-w_{2}\bigg(\frac{F_{1}^{2}}{2}+\frac{F_{2}^{2}%
}{2}\bigg)
\]
so that, in the absence of lobbying, the optimal policy for each regulatory dimension would be set at zero. The coefficients $w_{i}$ capture the cost of deviating from the optimal policy in each dimension, and are assumed to be large enough to always ensure an interior solution.\footnote{As it will become apparent below, a sufficient condition is $\min[w_{1}/a^{2},w_{2}/b^{2}%
]>9/2.$} The interpretation and microfoundation of the direct utility function are discussed in \cite{grossmanhelpmanbook}. 

\subsection{Analysis}

Firms first play the lobbying game with the regulator, when the policy vector $\mathbf{P}$ and the transfers are determined, and then they play the competition game, when quantities are set. We solve the game backwards.

In the last stage, standard calculations obtain%
\[
\pi_{i}=\frac{[A+aR+b(2F_{i}-F_{j})]}{9}^{2}.
\]

Lobbying over the common component impacts positively both firms, while the private component has opposing effects.

We now turn to the first stage. In order to show the differences between lobbying over the common and the private component, we analyze each case separately.

\subsubsection{Lobbying only over the common component \texorpdfstring{$R$ ($b=0$)}{R (b=0)}}

The policy maker selects $R$ to maximize%
\[
2(A+aR)^{2}/9-w_{1}R^{2}/2.
\]

When firms lobby over the common component, the common agency game becomes a public good contribution game with multiple equilibria. All equilibria lead to the same level of regulation and total lobbying spending. However, they may allocate spending differently to the two firms. The set of equilibria is as follows:

\begin{proposition}\label{prop:bis0}
When $b=0$, in a truthful equilibrium, regulation is given by:
\begin{equation}
R^{\ast}=\frac{4Ak_{R}/a}{9-4k_{R}}\label{Rduo}%
\end{equation}
where $k_{R}\equiv a^{2}/w_{1}$, and total contributions are:
\begin{equation}
\hat{t}_{1}+\hat{t}_{2}\geq w_{1}R^{\ast2}/2=\frac{8A^{2}k_{R}}{(9-4k_{R}%
)^{2}}.\label{ineqtotduo}%
\end{equation}
\end{proposition}

The transfers therefore reflect the policy given by (\ref{Rduo}). The comparative statics are sensible: policy and transfers are higher the larger the affected market (high $A$), and the higher $k_{R}$ is, that is, the more effective the policy is (high $a$), and the cheaper the social cost (low $w_{1}$). Notice in particular how transfers are convex in market size $A$.

\subsubsection{Lobbying only over the private components \texorpdfstring{$F_{i}$ ($a=0$)}{F (a=0)}}

The policy maker now selects $F_{i}$ to maximize%
\[
\frac{\lbrack A+b(2F_{1}-F_{2})]}{9}^{2}+\frac{[A+b(2F_{2}-F_{1})]}{9}%
^{2}-w_{2}\frac{F_{1}^{2}}{2}-w_{2}\frac{F_{2}^{2}}{2}.%
\]

This is no longer a common-interest game for the firms. Their interests go in opposite directions: a higher level of private regulation for one firm benefits that firm and hurts the other firm. In the lobbying phase, firms will jostle to supply favors to the policy-maker. Applying Corollary 1, we can find a closed-form expression for policy and transfers. Unlike the common-interest game, there is a unique truthful equilibrium game:

\begin{proposition}\label{prop:ais0}
When $a=0$, in a truthful equilibrium, regulation is given by:
\begin{equation}
F_{1}^{\ast}=F_{2}^{\ast}=\frac{2Ak_{F}/b}{9-2k_{F}}\label{Fduo}%
\end{equation}
where $k_{F}\equiv b^{2}/w_{2}$, and total contributions are:
\begin{equation}
\hat{t}_{1}+\hat{t}_{2}=\frac{36A^{2}k_{F}(5-2k_{F})}{(9-2k_{F})^{2}(9-10k_{F})}.\label{tm}%
\end{equation}
\end{proposition}

Comparative statics are similar to the previous case and omitted.

\subsection{The consequences of a merger}

Imagine now the two firms merge to a monopoly. The profit of the resulting merged firm, denoted as $m$, is%
\begin{align*}
\pi_{m}  & =\pi_{1}+\pi_{2}=p_{1}q_{1}+p_{2}q_{2}=\\
& (A+aR-Q)Q+b(F_{1}q_{1}+F_{2}q_{2}).
\end{align*}
We immediately derive a first result: The monopolist produces only the product with the highest private lobbying component.

To see this, by contradiction, imagine $F_{1}>F_{2}$ and $q_{2}>0.$ Then reduce $q_{2}$ by an amount $\delta$ and increase $q_{1}$ by the same amount so that $Q$ does not change. Profits then increase by $b\delta(F_{1}%
-F_{2})>0.$ It follows that the monopolist will want to decrease $q_{2}$ down to zero.\smallskip

From now onwards, we then take that the monopolist is interested in only one product (we take that to be Firm 1's product). After setting the optimal quantity, the equilibrium profits are
\[
\pi_{m}=\frac{(A+aR+bF_{1})}{4}^{2}.
\]
Turning now to the lobbying game, the policy maker selects the whole policy to maximize%
\[
\frac{(A+aR+bF_{1})}{4}^{2}-w_{1}\frac{R^{2}}{2}-w_{2}\frac{F_{1}^{2}}{2}%
\]
resulting in%
\begin{equation}
R^{m}=\frac{Ak_{R}/a}{2-k_{R}-k_{F}};\text{ }F_{1}^{m}=\frac{Ak_{F}/b}%
{2-k_{R}-k_{F}};\text{ }F_{2}^{m}=0.\label{Rm}%
\end{equation}

When it comes to the determination of the transfers, the merged firm just needs to compensate the regulator for the social loss $w_{1}R^{m2}%
/2-w_{2}F_{1}^{m2}/2,$ which after substitution amounts to
\begin{equation}
\hat{t}_{m}=\frac{A^{2}(k_{R}+k_{F})}{2(2-k_{R}-k_{F})^{2}}.\label{tm}%
\end{equation}

We are now in a position to state our final and main result.

\begin{proposition}\label{prop:merger}
A merger between the two firms can cause an increase or a decrease in regulation and in the total amount of lobbying spending by the firms to the policy-maker.
\end{proposition}

The proposition is proven by showing that in our two leading examples -- pure common good and pure private good -- the sign of the effect of a merger is opposite. 

The intuition is as follows. With $R$, lobbying produces a positive externality between firms because more spending by a firm benefits the other one. With $F$, it produces a negative externality because more spending from a firm hurts the other one (because it leads to higher quantity and thus lower price). From the industry perspective, in duopoly lobbying on $R$ is underprovided while lobbying on $F$ is overprovided. By merging, the firms solve the coordination problem and address the externality. For $R$, it means more lobbying. For $F$, it means less lobbying.\footnote{Given the way we modeled demand, with a monopoly, the remaining private component has the same effect as $R$ and, in the policy game, both $F$ and $G$ will be provided given the benefits that can be obtained from spreading lobbying resources between those two dimensions. 

Clearly, this result depends on our formulation, and it would be possible even to construct cases where the private component is eliminated altogether after a merger.} As for transfers, the key aspect for the reduction of lobbying transfers after the merger comes from the binding constraints in the transfer game. Before the merger, each firm has to compensate the policy maker a lot because the policy maker could otherwise set a ``bad'' individual component for that firm. After the merger, that threat is lost and total contributions do not need to be as high, but only reflect the cost to the policy maker. Although total lobbying may be lower, this does not imply that policy is necessarily better for consumers. 

At the current level of abstraction, our analysis cannot be used to draw welfare conclusions. The utility functions we use -- the firms' profit functions and the policy-maker's direct utility -- do not tell us how consumers are affected by changes in quantities, prices, and regulation. It is difficult to imagine a welfare function that applies to all markets and all types of regulation. By focusing only on $R$, one can interpret it as a tariff, which under standard assumptions hurts consumers. $R$ could also be interpreted as public investment in the enforcement of safety rules (increasing demand through an increase in trust), which could instead benefit both firms and consumers.  

\section{Empirical Overview}\label{sec:empoverview}

We now turn to measuring this set of theoretical ideas in a sample of real companies. We examine publicly-listed firms from 1999-2017 and their influence activity on the U.S. federal government. Among these firms, merging is a high-stakes strategic activity. As such, measuring the causal impact of mergers requires an identification approach. 

In this section we lay out our broad set of strategies and the panel data we use to execute it. A key conceptual tool is the notion of a \emph{composite treatment}. A composite treatment is a function of multiple inputs that interact to form a treatment. Recent literature in econometrics \citep{borusyak2020non} propose design-based theory and methods to handle composite treatments. These new methods specifically address empirical settings where some inputs to the composite treatment are highly endogenous, and other inputs may be influenced by quasi-random variation.

We adapt and apply these methods to studying mergers. In our data, a merger is a composite treatment that accepts two broad inputs: i) The merging parties and terms, and ii) the completion date. In the strategies we deploy below, we explicitly focus on ii) the \emph{timing of the merger} as the source of exogenous variation, holding the merging partners fixed. Although our data are from a non-experimental setting, the experimental equivalent is to hold fixed the merging parties, and randomly perturb the moments in time when the mergers are consummated. Because of our emphasis on timing, we use a set of panel data methods outlined below. 

By focusing on timing, we do not claim that firms' choices of merger partners are completely nonrandom. This is not a requirement of our identification strategy. Timing is simply one of many possible sources of variation that could be used to measure the effects of mergers. Future researchers may identify natural experiments in the choice of partners (or other aspects of merging).\footnote{Because these future designs would use different variation, they would potentially yield different local average treatment effects. Our theory section above offers some guidance about why treatments could vary.}

\subsection{Data Structure: Composite Firms}

Our approach to studying mergers uses a new unit of analysis called a \emph{composite firm}. Composite firms are clusters of multiple firms that eventually merge together by the end of the sample. For each \emph{component} firm (original, underlying firms), we can identify its \textit{composite} firm at the beginning of the sample (before the merger takes place). We can link each firm to composite (and siblings) for all periods in the sample, and leverage  \emph{within-composite firm variation} over time. Composite firms do not exist in standard merger databases, but can be assembled from the standard datasets about mergers and their timing. We developed the concept for this analysis. To our knowledge, our paper is the first to assemble the composite firm graph, study its evolution over time or use it for identification.

Appendix \ref{ComposFirmExample} presents a visualization of a multi-merger composite firm as a graph, and shows how we represent this firm in a regression-friendly panel matrix. Using the composite firm graph, we can observe the evolution of each composite at every point in our sample -- including when the underlying component firms are independent, while they merge, and after they are completely unified.

By representing merger activity through composite firms, we focus on exogenous variation in merger timing. The composite representation is particularly helpful in analyzing multi-merger firms. Mergers are relatively rare. However, among companies that \emph{do} merge with others in our sample, 42\% are involved in multiple mergers or acquisitions.\footnote{This number rises to 68\% if unlisted companies are included.} Multi-merger firms are particularly common among larger companies that may be the source of important political and/or economic influence. Composite firms with more than two components comprise 58\% of all lobbying spend.\footnote{This number rises to 83\% if  unlisted companies are included.} Such firms are often both targets and acquirers in the same sample. Appendix \ref{WhyCompFirms} describes why these present challenges both for representing the phenomena for identification and standard errors, and how our composite firm representation addresses the challenges. 

Our sample includes around 12K composites. These 12K composite firms are composed from over 15K \emph{component} firms in our original Compustat sample. Each of the 15K component firms has exactly one composite parent into which it is eventually merged. Many component firms never merge with any others; its composite parent is (essentially) itself. Using this panel of composite firms, we execute multiple identification strategies, all focused on the timing of mergers. 

\subsection{Regression Equation}

Our results come from estimating panel regressions of the following format: 
    \begin{equation}\label{Spec1}
    \begin{split}
        \sum_{f\in\mathcal{F}_{it}}y_{f,t} = & \beta_{0}+  \beta_{1}\text{MergerIndex}_{it} +\beta_{2}X_{it} +\delta_{i} + \gamma_{t} + \epsilon_{i}.
    \end{split}
    \end{equation}

$y_{f,t}$ represents political influence spending of firm $f$ at time $t$. We examine two measures of political influence (discussed in our data section): Federal lobbying spend and donations from political action committees. $\mathcal{F}_{it}$ represents the composite firm ownership partition for a composite firm $i$ at time $t$. As such, $\sum_{f\in\mathcal{F}_{it}}y_{f,t}$ represents the sum of all lobbying of all component firms in composite firm $i$ at time $t$. We include fixed effects for composite firms ($\delta_{i}$) and time periods ($\gamma_{t}$). Standard errors are clustered by composite firms. 

The coefficient of interest is $\beta_{1}$, the coefficient on the $\text{MergerIndex}_{it}$. In our main specification, we examine a simple count of the number of independent firms within each composite firm $i$ at time $t$. This decreases each time a merger occurs, and allows $\beta_{1}$ to be interpretable as the effect of a merger.\footnote{Our specification permits other measures of concentration as well, such as the Herfindahl-Hirschman Index (HHI), adapted to our application. Appendix \ref{app:HHI} implements this approach, and shows that our empirical results are qualitatively similar to using this alternative.} Because a merger corresponds to a \emph{decrease} in the number of firms within the bundle, a negative coefficient means that political spending \textit{increased} after the merger. 

Because mergers are endogenous, we examine several different approaches to identifying effect (outlined below). We vary controls $X_{it}$ in coordination with our identification strategies. Because of the potential importance of size, we control for total composite firm $revenue_{it}$ in all specifications.\footnote{In our notation, $revenue_{it}=\sum_{f\in\mathcal{F}_{it}}r_{f,t}$ where $r$ is revenue of each component firm.} We also use controls to increase precision of our main estimates, to report descriptive patterns of interest and as checks on the robustness of our findings \citep{altonji2005selection, oster2019unobservable}. In some specifications, we also control for trends by industry and other firm characteristics. 

For identification, we pursue two strategies. The first is a panel event study \citep{gentzkow2011effect, de2020two, freyaldenhoven2021visualization, goodman2021difference, athey2022design}. Our second is an exposure design, akin to a Bartik instrument \citep{bartik1991benefits, borusyak2020non, goldsmith2020bartik, breuer2021bartik}. In this approach, we develop an instrument for $\text{MergerIndex}_{it}$. Our instrument uses economy-wide shocks to the attractiveness of merging. Both designs are based on the timing of mergers. In order to explain our designs, we first describe the structure and sources of our data in the section below. 

\section{Data and Descriptive Statistics}\label{sec:data}

Our study of public firms from 1999-2017 combines data from four sources. This section describes these sources and summarizes key properties of the resulting dataset. We describe our sample and the data sources in detail below. In Appendix \ref{MergingData}, we describe how these data sources are merged together with identifiers. 


\subsection{Sample}

Our underlying sample consists of all firms present in the Compustat database from 1999-present. This includes publicly traded companies as well as private companies that are large enough to publicly disclose financial statements. This sample is limited in part by data availability. As discussed above, our empirical strategy requires pre-merger size data for all component firms. We use Compustat to obtain a sample of firms and key firm financial data including size (revenue) and industry (NAICS).\footnote{For each company, we use the first non-missing NAICS code} This sample is similar to those used in other studies of mergers between public firms.\footnote{See, for example, \cite{gaspar2005shareholder}, \cite{harford2011institutional}, \cite{bena2014corporate}.}

Our sample boundaries are also limited by the availability of political influence data. Detailed data on federal lobbying began only in 1999 following the Lobbying Disclosure Act (``LDA'') of 1995. LDA reports are required only once every half-year. As a result, half-years are the temporal unit of our panel, and we summarize all variables at the half-year level.\footnote{In 2007, a new disclosure law was adopted (``The Honest Leadership and Open Government Act'') requiring that lobbying disclosures take place twice as often (quarterly). Nonetheless, we continue our analysis on a half-year basis for consistency.} We include all firms that are available in Compustat for each half-year.

\subsection{Merger Data and the Composite Firm Graph}\label{dat:MergersComposites}

Our composite firm database uses Thomson Reuters' SDC Platinum database of acquisitions and mergers. SDC Platinum contains the universe of global M\&A transactions and is used in many academic papers about M\&As \citep{matvos2008cross, rossi2004cross, blonigen2016evidence}.\footnote{\cite{barnes2014evaluating} independently evaluate the SDC Platinum database and find positive results, particularly for the variables, time horizons and types of companies (larger) we analyze in this paper.  \cite{bollaert2015securities} evaluate other sources of merger data, including Zephyr ({\url{https://zephyr.bvdinfo.com/}}) and find positive results for SDC.} For each acquisition, SDC Platinum identifies the acquirer, target and dates associated with the merger.\footnote{The SDC dataset also includes other variables (such as the date of the merger announcement) as well as non-merger events such as rumored mergers. We do not use these in our analysis.} The date variables are particularly important in our analysis as they allow us to use pre-/post- variation in merger status. 

Using the methods in Appendix \ref{CompositeGraphProc}, we produce the composite firm graph. This procedure can be run for any time during our sample period. The procedure takes the above merger dataset and a date. For each underlying component firm, we identify a set of sibling firms who are connected through a merger or acquisition happening \emph{before the specified date}. This procedure is ``transitive'' in the sense that if Firm A is bought by Firm B, which is then purchased by Firm C -- then A is not only siblings with B, but also with C. Together, they form a composite firm which we can call ``ABC.'' We run the procedure using the final date of the sample. This assembles composites using all connections between firms at any point during our sample. We use this set of 12K composite firms as the $i$ variable in our $i\times t$ panel. 

The final step of this process is to measure the evolution of each composite firm over time. As discussed earlier, our identification strategy uses \emph{within composite firm} variation in concentration. To measure this, we run the procedure in Appendix \ref{CompositeGraphProc} for each half-year (the $t$ dimension of our panel) in our sample. This produces a dataset that connects each component firm $j$ to its eventual parent $i$, as well as to its intermediate parent $k$ at time $t$. The intermediate parent $k$ is a potentially smaller composite firm (i.e. collection of merged firms) that eventually merges into the main composite firm. Alternatively in cases towards the end of our sample, the intermediate parent $k$ is the final composite firm. 

Using these intermediate firms, we calculate the change in concentration over time. Our simplest measure of concentration is a count of the number of intermediate firms that still remain un-merged with each composite $i$ at each time $t$. This variable consists of integers that decrease by 1 with each successive merger.\footnote{On rare occasions when a firm merges with two firms within the same period, this number would decrease by two.}


\subsection{Political Influence Data} 

Our federal lobbying data comes from \textit{LobbyView},\footnote{\url{https://www.lobbyview.org/}} an NSF-funded project compiling federal lobbying data \citep{kim2017political,kim2018lobbyview}. These data have been used in several other papers.\footnote{See \cite{bombardini2020empirical, huneeus2018effects, ellis2018lobbies} for examples.} As discussed above, lobbying disclosures are required on a half-year basis (quarterly after 2008). The disclosures are made on forms that \textit{LobbyView} converts into structured, machine-readable data.\footnote{An example of a lobbying disclosure report can be viewed \href{https://soprweb.senate.gov/index.cfm?event=getFilingDetails&filingID=50767BDD-9687-4F63-BDC1-898E0EEAE624&filingTypeID=51}{here}.} Importantly, \textit{LobbyView} matches companies not only on its name, but also to a structured identifier that we can merge with our other data. 

For each company, \textit{LobbyView} contains disclosures for in-house lobbyists as well as lobbying performed by external firms hired by the company. Lobbying firms are required to identify their clients in these disclosures, so we can sum each company's in-house and outsourced lobbying. One limitation of this data is its handling of industry associations or coalitions. If a company donates money to an intermediary who hires lobbyists (such as an industry association or nonprofit), the intermediary's lobbying would be attributed to the intermediary. It cannot be traced back to the originating company/donor. This issue affects all research that uses lobbying data from the disclosure laws. 


Finally, we utilize data about campaign contributions. Our data about this outcome come from the Center for Responsive Politics' \emph{OpenSecrets} project.\footnote{\url{https://www.opensecrets.org/bulk-data/}} Other papers have used this data \citep{i2012revolving, bertrand2014whom}. Like \emph{LobbyView}, the \emph{OpenSecrets} project takes government disclosures and standardizes them into machine readable format. The \emph{OpenSecrets} process of standardization includes a greater level of manual review than \emph{LobbyView}. Coverage spans the 1998 electoral cycle to 2018. Campaign contributions include contributions from companies' PAC, as well as contributions by employees or owners of the organizations, as well as these individuals' family members. Before the Citizens United decision in 2010, companies could not directly donate to political campaigns. Afterwards, companies can donate directly to ``Super PACs'' (PACs with greater spending discretion), and these contributions are included in our dataset. 

\subsection{Summary Statistics}


Tables \ref{tab:DescStats}-\ref{tab:DescCorr} display summary statistics about our composite firms. Five broad patterns are evident. Although these patterns have been documented elsewhere in the literature, we mention these to set the context of our empirical application.

\begin{itemize}

\item[1.] {\bf Mergers among public companies are not uncommon}. 45\% of composite firms have been involved in a merger, although most of these mergers are acquisitions of small, unlisted companies. 10\% of our composite companies feature a merger between Compustat-listed companies.

\item[2.] {\bf Political influence is rare (per firm) but increasing over time}. 84\% of composite firms in our data have no lobbying, at any time during our sample, in any component firms. Similarly, 92\% of composite firms have no corporate PAC, for any composite firm, for any time during our sample. On the individual donor side, only 29\% of composites have at least one individual donor reported who listed one of the component firms as an employer. Spending on lobbying has grown over time in aggregate.

\item[3.] {\bf Firms spend a relatively small amount of revenue on political influence}. As described above, most firms' lobbying accounts for 0\% of revenue. Among those who do, the average amount is approximately one-hundredth of one percent. 

\item[4.] {\bf Firms spend more on lobbying than on campaign contributions}. This is true in aggregate, but also at the individual composite firm level. Of composite firms that spend at all on donations and lobbying, 90\% spend more on lobbying. 

\item[5.] {\bf Merging, revenue and political influence activity are correlated}. Large composite firms are more likely to lobby and have PACs and individual donors. They are also more likely to merge with another Compustat-listed firm and to have a longer lifespan. 


	\end{itemize}

Our descriptive tables present these patterns at the composite level, but we find the same patterns in our disaggregated dataset of individual component firms as well. Most component firms, even when viewed separately from their (eventual) merger siblings (where applicable) do not merge with other publicly listed firms or engage in political influence (\#1 and \#2) often or ever. Most underlying firms spend a relatively small amount of revenue on political influence (\#3), and spend more on lobbying than on campaign contributions (\#4). Component firms that merge are more likely to have high revenue and spend on politics.

The averages in Tables \ref{tab:DescStats}-\ref{tab:DescCorr} also highlight some important dimensions of heterogeneity. While most firms do not lobby, there is a sizable minority of firms who lobby a lot. Conditional on lobbying, the average composite firm spends over half of a million dollars on lobbying per year (\$670K) in our sample (median of \$56K/year). At the top of the distribution, there are firms that spend tens of million of dollars per year. As the raw correlations in Table \ref{tab:DescCorr} show, these firms tend to be the largest firms and are also more likely to engage in merger activity, which is the core question of our paper.

Other trends emerge along the time dimension. In the two decades of our sample, total lobbying spend steadily increased by \$67.2M per year on average. Among the firms that lobby in our sample, total lobbying spend increased by \$25.2M per year. This is an annual increase of \$3.6K per composite firm, or \$24.4K among firms who lobby at all. Among firms lobbying at all, the median lobby spend increased by 2.5 times, from \$80K in 1999 to \$200K in 2017, a large increase. Also during this period, the number of firms at any cross-section of our sample decreased by less than 1\% per year. The reduction in publicly traded companies has been documented in other studies \citep{grullon2015disappearance, doidge2017us}. The proportion of these firms in our sample that were lobbying at any time increased very slightly over time (by less than 1\% per year).\footnote{One reason for this is our composite firm level of analysis. If a company does not lobby but its future merging partner does, we count both companies as part of the same composite firm and are coded as lobbying. Similarly, when two lobbying companies merge and continue lobbying, we do not treat this as a reduction in the number of firms lobbying.}


\section{Panel Event Study}\label{sec:panelevent}

Panel event studies are a type of econometric model studied by  \cite{de2020two, freyaldenhoven2021visualization, goodman2021difference,  athey2022design}. In this approach, estimation of Equation (\ref{Spec1}) is straightforward (i.e., there is no instrument or first stage). In this setup, mergers are endogenous, but we assume they depend on fixed (or slow-moving) variables whose trends we control for. The consummation of the merger creates a sharp discontinuity in the firms' ability to coordinate externalities. 

The threat to identification in this strategy comes from a potential unobserved confound $C_{it}$. $C_{it}$ can include potentially unobserved time-specific factors for each composite firm, as well as an idiosyncratic component i.e.,  $C_{it}=\lambda_{i}'F_{t} + \xi\eta_{it}$. \cite{freyaldenhoven2021visualization} notes that Equation (\ref{Spec1}) is identified with two-way fixed effects model, as long as $C_{it}$ is low-dimensional and $F_{t}=0$. In our setting, a confound would violate this criteria if it affects political influence activities through a non-merger mechanism, and would coincide with the merger event. 

To complement this approach, we also add unit-specific, time-varying controls that may capture such confounds. In particular, we include a measure of firm size (revenue) and allow for industry-specific trends at a narrow category (NAICS5). We also include firm-specific political cycle effects,\footnote{Our firm-specific political cycle controls would capture the possiblity that ``Walmart tends to spend a lot in the midterms,'' or ``Boeing spends a lot during the presidential election years'' and etc. To implement this, we codify each half-year in our sample based on its timing within a four year (eight half-year) political cycle between presidential elections. The main effect of political cycles is absorbed by our half-year fixed effects. We then interact these cycle indicators with firm identifiers to produce firm-specific political cycle effects.} as well as controls for differential revenue effects depending on the number of mergers during the sample. The identification assumption is that the timing of the mergers, after conditioning on these other factors, comes from idiosyncratic shocks that are unrelated to the returns of political spending. 

A challenge that is unaddressed by this specification is the possibility of pre-merger increases in lobbying activity. Firms could initiate this form of pre-merger lobbying to influence the merger's review by regulators. Alternatively, firms may anticipate a positive review, and begin coordinating and integrating lobbying activity before the official merger date. Note that such pre-merger activity would bias the ``control'' period upwards, resulting in a smaller difference coming from the merger. The resulting bias is likely to work against finding a positive effect by inflating the pre-merger levels. We  address this potential with an additional specification controlling for anticipation effects (the results are summarized in the next section and reported in Appendix \ref{app:anticip}). 


\paragraph{Results.} Table \ref{tab:MainPoliSpend} shows results on lobbying and PAC donations using our main specification in Equation (\ref{Spec1}). Columns 1 and 3 include two-way fixed effects and revenue controls. Columns 2 and 4 contain the additional controls described above. 

In all of our specifications, results point in the same direction: Greater concentration increases composite firms' spend on political influence activities (both lobbying spend and PAC spend). Our results suggest the average merger increases lobbying spend by \$140,000 per year (column 2), while the impact on PAC donations amounts to almost \$8,000 per year (column 4). Results are robust to using HHI instead of the number of component firms as an index of concentration (see Appendix \ref{app:HHI}).

\begin{figure}[H]
\caption{\textbf{Lobby Spending: Event Study Plots} \label{fig:PanelEventStudyPlotLobby}}\center
\includegraphics[height=15cm]{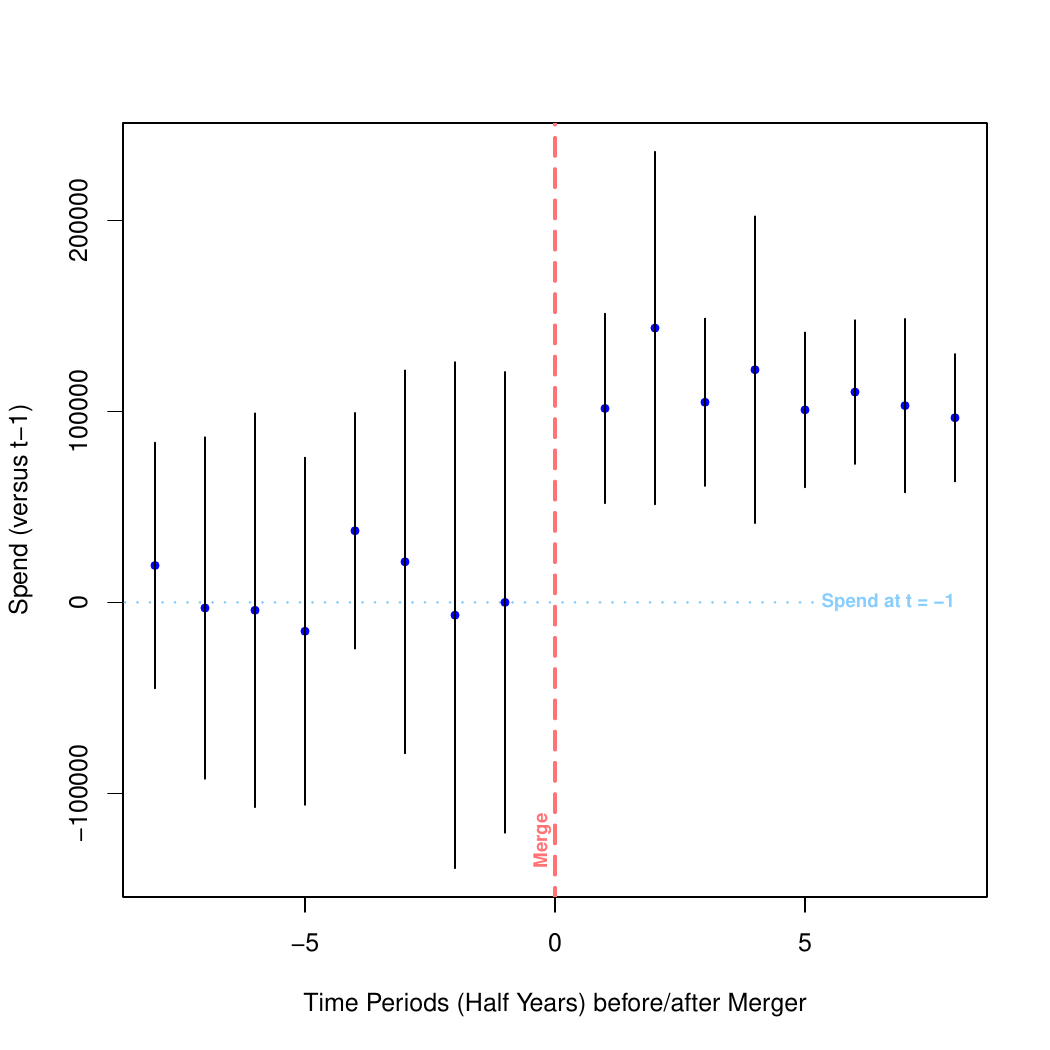}
    \begin{footnotesize}
    \begin{singlespace}
    \flushleft \vspace*{-5mm}
	\textbf{Notes}: This figure shows an event study plot showing displaying differences before and after the merger (window length = eight half years before/after), using our event study design. Each point bar represents the cumulative effect of the merger on per- half year spend.
	\end{singlespace}
	\end{footnotesize}
\end{figure}

\begin{figure}[H]
\caption{\textbf{PAC Donations: Event Study Plots} \label{fig:PanelEventStudyPlotPac}}\center
\includegraphics[height=15cm]{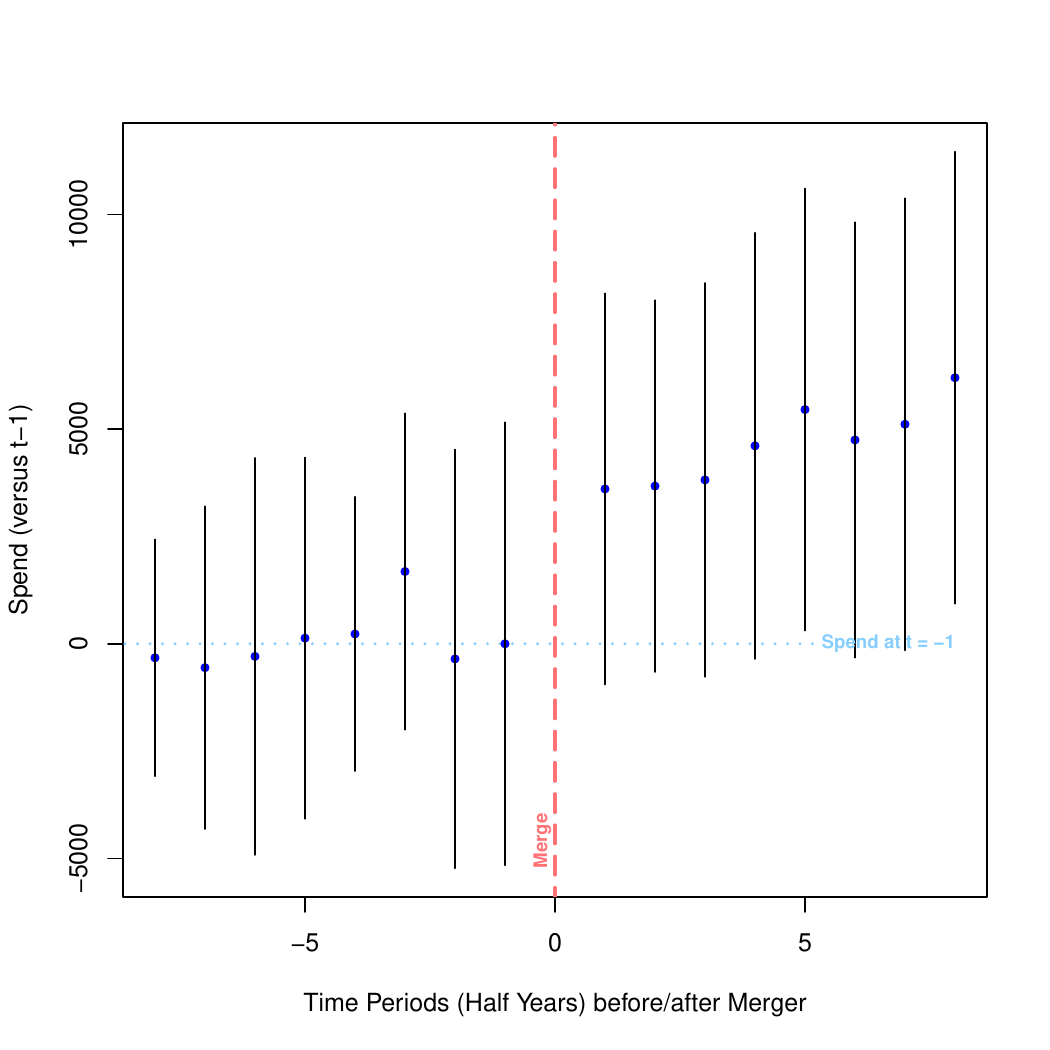}
    \begin{footnotesize}
    \begin{singlespace}
    \flushleft \vspace*{-5mm}
	\textbf{Notes}: This figure shows an event study plot displaying spending differences before and after the merger (window length = eight half years before/after), using our event study design. Each point bar represents the cumulative effect of the merger on per- half year spend.
	\end{singlespace}
	\end{footnotesize}
\end{figure}

To visualize these effects, Figures \ref{fig:PanelEventStudyPlotLobby} and \ref{fig:PanelEventStudyPlotPac} display event study plots. Each point bar represents the cumulative effect of the merger on per-period spend at each period of time.\footnote{These plots include a window of 8 periods, or four years, on either side of the merger. In some approaches to event study plots, coefficients are estimated to place additional bars on the plot that aggregate for all pre- and post- window observations. We have not estimated these coefficients as they significantly decrease our sample size.} Although some data points are estimated noisily, the broad pre/post effects are visible in the plot. 

We also probe the robustness of our results to pre-merger anticipation effects (See Appendix \ref{app:anticip}). One could imagine that merging firms may engage in lobbying activities to get the merger approved. However, our data show no evidence of increased lobbying or campaign spending in the six months that precede the merger.

This null result is consistent with the observation that in the period under consideration the US antitrust authorities scrutinized a small proportion of mergers \citep{wu2018curse}. Between 2010 and 2019, the Federal Trade Commission and the Department of Justice issued ``Second Requests'' to between 2.2\% and 3.9\% of transactions depending on the year \citep{hartscottrodino}. This means that in each of those years over 95\% of proposed mergers were approved within 30 days with no additional information requests.

\paragraph{Heterogeneity: Size and Similarity.} Our specification allows us to examine heterogeneity across different types of firms. Our theory features two aspects in particular. First, it is a theory of horizontal mergers of similar firms. Second, our theory intuitively should apply particularly to ``large'' firms, especially if there are fixed costs associated to lobbying. 

We can operationalize these concepts using our data. For size, we use revenue. We sum all revenue across the entire sample for each composite firm, and examine companies above and below the median.\footnote{Although this splits our composite firms in half, it does not split our entire panel in half because the large firms have more observations, possibly because of survivorship bias.} In Table \ref{tab:MainPoliSpendSize}, we find that although mergers broadly increase lobbying spend across both sets of firms, the effects on large firms are bigger. 


Our theory also suggests that a merger of more closely-related firms would have a bigger effect. Such firms are more likely to have common, overlapping interests. To measure close vs. distant mergers, we use data about the industry categorizations of  component firms (measured by NAICS codes). For each composite firm, we measure the number of unique NAICS codes at the beginning of the sample. Composite firms with a high number of unique NAICS codes represent firms that merge across industries (distant), while those with few unique NAICS codes represent within industry mergers. 

Tables \ref{tab:CloseVsDistant} shows our close-vs-distant results. We interact our $\text{MergerIndex}_{it}$ variable with our measure of distance. Our findings suggest that mergers among more distant firms have a lower overall increase in lobbying. The effect on political activity is instead higher when the merging firms are within the same industry.

\section{Differential Exposure Design}\label{sec:exposiv}

Our second approach to identification is an exposure design  \citep{borusyak2020non, goldsmith2020bartik, breuer2021bartik}. The idea in these designs is that units are affected by shocks, but they have differential exposure to these shocks. In an influential paper developing this strategy, \cite{bartik1991benefits} examined how employment growth affects wage growth. Because employment growth is endogenous, the authors developed an instrument. The instrument exploited the idea that economy-wide demand shocks have idiosyncratic effects in local markets. These shocks varied systematically according to the pre-shock characteristics of the local market. 

In this section we pursue a similar strategy to study mergers. A long-noticed fact about mergers is that they arrive in waves \citep{nelson1959merger, gort1969economic, weston1990mergers}. These waves span multiple sectors \citep{maksimovic2013private}, and have several underlying causes including macroeconomic shocks \citep{maksimovic2001market, rhodes2004market}, regulatory and technology shocks \citep{mitchell1996impact, harford2005drives}, uncertainty \citep{toxvaerd2008strategic, bonaime2018does}, connections between industries \citep{ahern2014importance}, and even envy among CEOs \citep{goel2010envious} and management fads \citep{haleblian2012exploring}. 


We utilize economy-wide pro-merger shocks at different times to construct a time-varying instrument similar to the \cite{bartik1991benefits} instrument. At various times during our sample, mergers have been particularly popular (or unpopular) compared to the overall trends. We measure these shocks, and interact them with measurements of a firm (or industry's) exposure to these shocks. As we show later, our instrument has a strong first stage. 

\sloppy 
\subsection{Implementation} To implement this design we again use Equation (\ref{Spec1}), but develop an instrument for the key measure of concentration. The unit of observation in this regression is \{half year\}$\times$\{composite firm\}. The instrumented variable is $\text{MergerIndex}_{it}$, which measures how concentrated composite firm $i$ is at time $t$. As is common for exposure designs, our instrument is a product of two terms. 

\paragraph{Merger Wave Term (Time-Varying).} The first term is the average $\text{MergerIndex}$ for other firms in the same period, excluding the focal firm \emph{and all other firms' in the focal firm's industry.} The first term can be written as: 
\begin{equation}
    W_{it}=\frac{\sum_{j: S_{i}\neq S_{j}}MergerIndex_{j,t}}{N_{S_{i}\neq S_{j}}}
\end{equation}
where $S_{i}$ and $S_{j}$ represent the industries of composite firms $i$ and $j$. $W_{it}$ captures the time-varying merger waves; in periods with high concentration due to economy-wide shifts in concentration, $W_{it}$ will be high. 

As is typical in exposure designs, we measure these shocks using a ``leave-one-out'' average of changes in the same period. We go beyond this and leave out the entire industry of  each focal observation. By leaving out the entire industry, our goal is to ensure that we measure shocks arising from economy-wide trends and that are not part of the endogenous dynamics among close competitors. We define the focal industry broadly by using the top level NAICS category for each composite firm in its initial period.\footnote{NAICS classifications for composite firms are calculated for each period by summing the revenue in each NAICS category, and selecting the NAICS code with the most revenue.} As a result, the value of $W_{it}$ differs not only over time, but also across observations within the same time period. However, the main purpose of $W_{it}$ is to capture time-varying shocks to the entire sample. Because merger waves are indeed economy wide (in our sample and in others), shocks between different industries during the same time period are correlated. 


\paragraph{Exposure Term (Unit-Varying).} The second term is a cross-sectional feature of each composite firm at period zero. It represents the firm's exposure to merger waves. We call this term $K_{i0}$. This term already appears in Equation (\ref{Spec1}) as part of the composite firm fixed effects; it enters our IV strategy again when we create an instrument for $\text{MergerIndex}_{it}$ using the product of $K_{i0}$ and $W_{it}$. 

We examine several possible implementations of $K_{i0}$ for robustness. Our main exposure term we call $N_{i0}$, or the total number of component (member) firms inside each composite firm in its initial period. Defined this way, ``large'' composite firms (high $N_{i0}$) are more exposed to shocks; as there are more member firms who could merge together and increase the $\text{MergerIndex}_{it}$ for this composite firm. As a robustness check, we also implement $K_{i0}$ as the \emph{average} of $N_{i0}$ for all firms inside the same NAICS industry. In this representation, entire industries (rather than particular firms) have a greater or lower exposure to merger waves. 

Either way, high $K_{i0}$ indicates a high propensity to merge in the overall sample period. However, the specifications of $K_{i0}$ say nothing about the \textit{timing} of mergers, only about the overall propensity over the sample period. The timing could be anything, for example, a high exposure ($K_{i0}$) firm could be completely unresponsive to merger waves by (for example) doing all mergers in the first period and remaining inactive for the rest of the sample. There are many ways for a high $K_{i0}$ firm to avoid complying with merger waves. We integrate the timing aspect into the other term in the instrument (the wave term $W_{it}$, described above). 

We now have the main components of our instrument. Our instrument is  $Z_{it}=W_{it}K_{i0}$, the product of the wave term ($W_{it}$) and the exposure term ($K_{i0}$). Because Bartik-like instruments are products, researchers typically argue that one (or both) elements are exogenous \citep{goldsmith2020bartik}. Consistent with our discussion above, we portray the time-varying shocks as exogenous, and regard the identity of merging partners (and thus the level of exposure) as endogenous. As in our earlier design, identification comes from merger timing. 

We use this $Z_{it}$ to instrument the $\text{MergerIndex}_{it}$ term in Equation (\ref{Spec1}) by using the following first stage regression: 
\begin{equation}\label{BartikFS}
    \begin{split}
        \text{MergerIndex}_{it} = & \lambda_{0}+ \lambda_{1}Z_{it} +\lambda_{2}X_{it} +\zeta_{i} + \tau_{t} + \eta_{i}.
    \end{split}
    \end{equation}
This is the same regression as Equation (\ref{Spec1}), but the dependent variable is now $\text{MergerIndex}_{it}$, and the main independent variable is now our instrument $Z_{it}$. The other terms are the same but given separate names; the coefficients are now $\lambda$s, the error term is $\eta$, $\zeta_{i}$ are composite firm fixed effects and $\tau_{t}$ are time period fixed effects. Diagnostics on the instruments (correlations tests, compliers and instrument strengths) are performed in Appendix \ref{app:BartikDiagnostic}.


\subsection{Results}

Table \ref{tab:ExposureDesign} shows results on lobbying and PAC spend using our exposure IV specification. Panel A contains our first implementation of $K_{i0}$, and Panel $B$ contains the second. As with our earlier results, our specifications suggest that greater concentration increases composite firms' spend on political influence activities (both lobbying spend and PAC spend). 

Our results in this design are in the same order of magnitude as the panel event study, although slightly larger (and also with larger standard deviations). This is consistent with our analysis in Appendix \ref{app:BartikDiagnostic}, showing that larger firms were more likely to be compliers to our instrument. The average merger identified by this design increases lobbying spend over \$200K per year (columns 1 \& 2), while the impact on PAC donations is around \$10K per year (columns 3 \& 4). 

In both of our empirical designs, our merger coefficients aggregate across lots of heterogeneity. Presumably some of the mergers have a more rivalrous quality. The results above speak to aggregated averages (and in the exposure design, a local average treatment effect). The positive sign of these coefficients suggests that the typical set of merging firms have overlapping, public good-like regulatory interests. Our theory model shows why this would cause firms to increase spend, as the positive externalities of lobbying become  internalized.


\section{Firm-Level Political Risk}\label{sec:altexplan}

In our theory section, a merger helps coordinate the positive externalities of lobbying for a common cause. However, another mechanism could also produce this increase: After a merger, regulators could increase scrutiny as a result of negative attention from third parties. Because of this attention, the merged entity could increase political spending --- not because of coordinated externalities, but in response to a more adversarial environment. 

To investigate this possibility, we examine measures of \emph{firm-level political risk}. If the political environment became more negative after a merger, then we may expect exposure to political risk to increase after the merger. A highly-cited paper by \cite{hassan2019firm} develops an empirical strategy for measuring firm-level political risk over time. The approach uses text-mining methods to quantify ``[T]he share of [a firm's] quarterly earnings conference calls that they devote to political risks.'' We use the measures from this paper as the outcome variables in our panel specifications above. 

The \cite{hassan2019firm} metrics not only contain an overall measure of firm-level risk, but additional detailed data about the \emph{type and direction} of political risk. The primary measure in \cite{hassan2019firm} is the overall level of political risk. However, they also score the sentiment of the discussions. Higher sentiment indicates more positive discussion. In addition, the data contains detailed breakdowns about the level of political risk across eight topics: Economic policy \& budget, environment, trade, institutions \& political process, health, security \& defense, tax policy, and technology \& infrastructure. Our main results include the economic policy \& budget variable as an outcome, but we include the full set of categories in the appendix. 

Political risk measures are available only for the subset of firms that have regular investor calls. Appendix \ref{app:investorcalls} contains descriptive statistics for firms that are in our investor call sample (compared to ones that are not), and other details of how we integrated this data into our composite firm panel. Our panel of composite firms that use investor calls is about one third of the size of the sample as a whole. Firms with regular investor calls are generally larger and more politically active. 

Table \ref{tab:PolRisk} contains our results using both our panel event study (Panel A) and exposure designs (Panel B). For ease of interpretation, we normalize all measures of political risk. In Columns 1 and 2, we replicate our main results on lobbying and PAC spending on the subsample. Our results on this subsample have the same direction and size as our main results -- although in some cases less precise, partly as a result of the smaller sample size (31\% of the main sample). 

The remaining columns show the effect of mergers on political risk, particularly risks around economic policy. We find no evidence of higher political risk after a merger (in any specification). Our estimates generally fail to reject zero, with standard errors small enough to rule out large effects. In one case, we obtain statistically significant results in the opposite direction: Political sentiment becomes more positive after the merger (although the size of this effect is small). Table \ref{tab:PolRiskAll} contains all measures of political risk,\footnote{In total we study ten measures of polical risk. Trade policy is one area where we do find a small statistically significant differences in risks after mergers.} and Appendix \ref{app:HHI} contains HHI versions. 

\section{Conclusion}\label{sec:conclusion}

Our paper has tried to contribute to the lively debate on the increase in industry concentration and changes in business dynamics \citep{philippon2019great,de2020rise,dube2020monopsony}, as well as its causes and policy implications \citep{autor2020fall,berry2019increasing,grullon2019us,azar2020labor,dube2020monopsony}.

We contribute to this discussion by adding an additional element (political influence) and studying how firms vie to get political power both in theory and in the data. Our theoretical model takes a standard model of competition, and extends it to include regulatory variables. While our data come from a developed economy within a democratic state, our model is agnostic about the form of government (or the level of development). In countries with less democratic accountability, some of the forces in our model could be stronger or weaker. State capture by business interests is an issue appearing in development economics \citep{canen2022political}. 

Our data from the U.S. suggests that firms increase lobbying after mergers. This pattern survives a number of robustness checks and alternative stories. The association is stronger for large firms, and for firms in the same industry. 

We see our set of findings not as conclusive, but we hope it is a starting point for richer and deeper analyses of the political effect of mergers. By focusing on specific industries, future research could explore the link between lobbying activity and government regulation. When a merger occurs, which policies is the additional influence activity directed toward? In turn, how do those policies affect the firms' profit and consumer welfare in that industry?

These findings do not dispute the benefits of many forms of regulation to consumers (e.g., safety or environmental reasons), or that mergers can sometime increase efficiencies. However, corporate control of regulations could be used to erect barriers to entry or otherwise protect incumbents' market power. This would constitute another form of consumer harm, but one delivered through the channel of regulation rather than price, quantity or innovation. 

Investigating this consumer harm is beyond the scope of this paper, but it is a natural avenue for future research. 


\singlespacing
\clearpage \putbib 

\clearpage 
\section*{Tables and Figures}
\vspace{-2mm}

\pagebreak
\FloatBarrier

\begin{table}[H]\centering
\caption{\textbf{Descriptive Statistics: All Composite Firms} \label{tab:DescStats}}

\resizebox{\textwidth}{!}{{
\def\sym#1{\ifmmode^{#1}\else\(^{#1}\)\fi}
\begin{tabular}{l*{1}{ccccccc}}
\hline\hline
                    &        Mean&     Std.Dev&         Min&         P25&         P50&         P75&         Max\\
\hline
Years in Sample     &        8.78&        6.44&        0.50&        3.00&        6.50&       14.50&       19.00\\
Avg Revenue (\$10M, per Half Year)&       62.95&      392.77&        0.00&        0.01&        1.87&       16.15&    18359.17\\
\hline Lobby Spend (\$1K, per Half Year)&       54.09&      558.41&        0.00&        0.00&        0.00&        0.00&    40365.12\\
Lobbied at all (per Half Year)&        0.08&        0.23&        0.00&        0.00&        0.00&        0.00&        1.00\\
In-House Lobby Spend (\$1K, per Half Year)&       36.33&      458.92&        0.00&        0.00&        0.00&        0.00&    37828.85\\
Lobbying Intermediary Spend (\$1K, per Half Year)&       17.76&      139.79&        0.00&        0.00&        0.00&        0.00&     7182.46\\
Lobbied at all (ever)&        0.16&        0.37&        0.00&        0.00&        0.00&        0.00&        1.00\\
\hline PAC Donations (\$1K, per Half Year)&        2.26&       25.03&       -0.12&        0.00&        0.00&        0.00&     1903.46\\
PAC Donations$>0$ (per Half Year)&        0.05&        0.19&        0.00&        0.00&        0.00&        0.00&        1.00\\
PAC Donations$>0$ (Ever)&        0.08&        0.27&        0.00&        0.00&        0.00&        0.00&        1.00\\
\hline Individual Donations (\$1K, per Half Year)&        0.59&        3.92&       -1.75&        0.00&        0.00&        0.01&      157.41\\
Individual Donations$>0$ (per Half Year)&        0.06&        0.14&        0.00&        0.00&        0.00&        0.05&        1.00\\
Individual Donations$>0$ (Ever)&        0.29&        0.45&        0.00&        0.00&        0.00&        1.00&        1.00\\
\hline Individual + PAC (\$1K, per Half Year)&        2.86&       27.16&       -1.75&        0.00&        0.00&        0.02&     2011.94\\
Individual + PAC$>0$ (per Half Year)&        0.09&        0.21&        0.00&        0.00&        0.00&        0.05&        1.00\\
Individual + PAC$>0$ (Ever)&        0.30&        0.46&        0.00&        0.00&        0.00&        1.00&        1.00\\
\hline Ever M\&A    &        0.10&        0.31&        0.00&        0.00&        0.00&        0.00&        1.00\\
\# of Component Firms&        1.24&        1.23&        1.00&        1.00&        1.00&        1.00&       39.00\\
\hline\hline
\end{tabular}
}
}

    \flushleft 
    \begin{small}
    \begin{singlespace}
	\textbf{Notes}: This table displays simple summary statistics for all composite firms and all periods in our sample. Our panel dataset is described in Section \ref{sec:data}, and  composite firms are defined at the beginning of Section \ref{sec:empoverview}. 
	\end{singlespace}
	\end{small}
\end{table}

\begin{table}[htbp]\centering
\caption{\textbf{Descriptive Statistics: Firms Who Lobby} \label{tab:DescStatsLobbyFirms}}

\resizebox{\textwidth}{!}{{
\def\sym#1{\ifmmode^{#1}\else\(^{#1}\)\fi}
\begin{tabular}{l*{1}{ccccccc}}
\hline\hline
                    &        Mean&     Std.Dev&         Min&         P25&         P50&         P75&         Max\\
\hline
Years in Sample     &       14.36&        5.78&        0.50&        9.50&       18.50&       19.00&       19.00\\
Avg Revenue (\$10M, per Half Year)&      274.98&      902.06&        0.00&        2.84&       38.04&      164.74&    18359.17\\
\hline Lobby Spend (\$1K, per Half Year)&      335.09&     1355.80&        0.16&        7.11&       28.96&      138.00&    40365.12\\
Lobbied at all (per Half Year)&        0.50&        0.35&        0.03&        0.16&        0.45&        0.86&        1.00\\
In-House Lobby Spend (\$1K, per Half Year)&      225.06&     1123.69&        0.00&        0.00&        0.00&       40.92&    37828.85\\
Lobbying Intermediary Spend (\$1K, per Half Year)&      110.03&      333.07&        0.00&        6.05&       23.34&       78.42&     7182.46\\
Lobbied at all (ever)&        1.00&        0.00&        1.00&        1.00&        1.00&        1.00&        1.00\\
\hline PAC Donations (\$1K, per Half Year)&       13.05&       60.22&        0.00&        0.00&        0.00&        4.57&     1903.46\\
PAC Donations$>0$ (per Half Year)&        0.25&        0.37&        0.00&        0.00&        0.00&        0.50&        1.00\\
PAC Donations$>0$ (Ever)&        0.38&        0.48&        0.00&        0.00&        0.00&        1.00&        1.00\\
\hline Individual Donations (\$1K, per Half Year)&        2.74&        8.89&       -1.75&        0.00&        0.13&        1.61&      157.41\\
Individual Donations$>0$ (per Half Year)&        0.19&        0.20&        0.00&        0.00&        0.13&        0.32&        1.00\\
Individual Donations$>0$ (Ever)&        0.71&        0.45&        0.00&        0.00&        1.00&        1.00&        1.00\\
\hline Individual + PAC (\$1K, per Half Year)&       15.78&       65.12&       -1.75&        0.00&        0.48&        7.63&     2011.94\\
Individual + PAC$>0$ (per Half Year)&        0.34&        0.36&        0.00&        0.00&        0.20&        0.62&        1.00\\
Individual + PAC$>0$ (Ever)&        0.74&        0.44&        0.00&        0.00&        1.00&        1.00&        1.00\\
\hline Ever M\&A    &        0.32&        0.47&        0.00&        0.00&        0.00&        1.00&        1.00\\
\# of Component Firms&        1.99&        2.69&        1.00&        1.00&        1.00&        2.00&       39.00\\
\hline\hline
\end{tabular}
}
}

    \flushleft 
    \begin{small}
    \begin{singlespace}
	\textbf{Notes}: This table displays simple summary statistics for all composite firms in our sample that lobby in at least one period. Our panel dataset is described in Section \ref{sec:data}, and  composite firms are defined at the beginning of Section \ref{sec:empoverview}. 
	\end{singlespace}
	\end{small}
\end{table}

\begin{table}[htbp]\centering
\caption{\textbf{Merged vs Non-Merging Composite Firms: Differences in Means} \label{tab:DescMergedNot}}

\resizebox{\textwidth}{!}{\begin{tabular}{l*{6}{c}}
                    &\Centerstack{Never\!Merged}&\Centerstack{Merged}&  Difference   \\
\hline
Years in Sample     &        7.98&       15.67&       -7.69***\\
Avg Revenue (\$10M, per Half Year)&       35.18&      300.14&     -264.97***\\
\hline Lobby Spend (\$1K, per Half Year)&       21.29&      334.31&     -313.02***\\
Lobbied at all (per Half Year)&        0.06&        0.30&       -0.24***\\
Lobbied at all (ever)&        0.12&        0.50&       -0.38***\\
In-House Lobby Spend (\$1K, per Half Year)&       12.46&      240.20&     -227.73***\\
Lobbying Intermediary Spend (\$1K, per Half Year)&        8.82&       94.11&      -85.29***\\
\hline PAC Donations (\$1K, per Half Year)&        0.82&       14.58&      -13.76***\\
PAC Donations$>0$ (per Half Year)&        0.03&        0.22&       -0.19***\\
PAC Donations$>0$ (Ever)&        0.05&        0.34&       -0.30***\\
\hline Individual Donations (\$1K, per Half Year)&        0.30&        3.08&       -2.77***\\
Individual Donations$>0$ (per Half Year)&        0.04&        0.20&       -0.16***\\
Individual Donations$>0$ (Ever)&        0.24&        0.78&       -0.54***\\
\hline Individual + PAC (\$1K, per Half Year)&        1.12&       17.66&      -16.53***\\
Individual + PAC$>0$ (per Half Year)&        0.06&        0.32&       -0.26***\\
Individual + PAC$>0$ (Ever)&        0.24&        0.79&       -0.55***\\
\# of Component Firms&        1.00&        3.33&       -2.33***\\
\hline\end{tabular}
}

    \flushleft 
    \begin{small}
    \begin{singlespace}
	\textbf{Notes}: This table displays average differences between composite firms that merge and composite firms that do not. Our dataset is described in Section \ref{sec:data}, and  composite firms are defined at the beginning of Section \ref{sec:empoverview}. 
	\end{singlespace}
	\end{small}
\end{table}

\begin{table}[htbp]\centering
\caption{\textbf{Descriptive Statistics: Correlations} \label{tab:DescCorr}}

\resizebox{.9\textwidth}{!}{\begin{tabular}{c*{7}{c}}
                &\multicolumn{6}{c}{}                                                         \\
                &\Centerstack{Years}   &\Centerstack{Revenue}   &\Centerstack{Lobby}   &\Centerstack{PAC}   &\Centerstack{Individual}   &\Centerstack{Ever\!Merged}   \\
\hline
\Centerstack{Years}&        1   &            &            &            &            &            \\
\Centerstack{Revenue}&     0.16***&        1   &            &            &            &            \\
\Centerstack{Lobby}&     0.13***&     0.49***&        1   &            &            &            \\
\Centerstack{PAC}&     0.12***&     0.49***&     0.84***&        1   &            &            \\
\Centerstack{Individual}&     0.18***&     0.42***&     0.49***&     0.49***&        1   &            \\
\Centerstack{Ever\!Merged}&     0.37***&     0.21***&     0.17***&     0.17***&     0.22***&        1   \\
\hline\end{tabular}
}

    \flushleft 
    \begin{footnotesize}
    \begin{singlespace}
	\textbf{Notes}: This table displays raw correlations between some of the key variables in our analysis. Our panel dataset is described in Section \ref{sec:data}, and composite firms are defined at the beginning of Section \ref{sec:empoverview}. 
	\end{singlespace}
	\end{footnotesize}
\end{table}

\begin{table}[htbp]\centering
\caption{\textbf{Results, Panel Event Study} \label{tab:MainPoliSpend}}

\begin{tabular}{l*{14}{c}}
                    &\multicolumn{1}{c}{(1)}&\multicolumn{1}{c}{(2)}&\multicolumn{1}{c}{(3)}&\multicolumn{1}{c}{(4)}\\
                    &\multicolumn{1}{c}{\Centerstack{Lobby\!Amount}}&\multicolumn{1}{c}{\Centerstack{Lobby\!Amount}}&\multicolumn{1}{c}{\Centerstack{PAC\!Contribs}}&\multicolumn{1}{c}{\Centerstack{PAC\!Contribs}}\\
\hline
\# Component Firms  &     -74,286** &     -68,934** &      -4,470*  &      -3,898   \\
                    &    (33,691)   &    (28,188)   &     (2,382)   &     (2,514)   \\
\hline
Additional Controls &               &           Y   &               &           Y   \\
\hline Observations &     223,043   &     223,022   &     223,043   &     223,022   \\
\(R^{2}\)           &         .79   &         .83   &         .32   &         .47   \\
\end{tabular}

    \flushleft 
    \begin{small}
    \begin{singlespace}
	\textbf{Notes}: This table shows results on lobbying and PAC donations using our panel event study specification in described in Section \ref{sec:panelevent}. 
	\end{singlespace}
	\end{small}
\end{table}

\begin{table}[htbp]\centering
\caption{\textbf{Heterogeneity by Firm Size (Panel Event Study)} \label{tab:MainPoliSpendSize}}

\begin{tabular}{l*{14}{c}}
                    &\multicolumn{1}{c}{(1)}&\multicolumn{1}{c}{(2)}&\multicolumn{1}{c}{(3)}&\multicolumn{1}{c}{(4)}\\
                    &\multicolumn{1}{c}{\Centerstack{Lobby\!Amount}}&\multicolumn{1}{c}{\Centerstack{Lobby\!Amount}}&\multicolumn{1}{c}{\Centerstack{PAC\!Contribs}}&\multicolumn{1}{c}{\Centerstack{PAC\!Contribs}}\\
\hline
\# Component Firms  &     -15,835   &     -66,208** &        -823   &      -3,788   \\
                    &    (17,269)   &    (28,513)   &     (1,107)   &     (2,513)   \\
\hline
Additional Controls &           Y   &           Y   &           Y   &           Y   \\
\hline Sample       &\Centerstack{Below\!Median\!Revenue}   &\Centerstack{Above\!Median\!Revenue}   &\Centerstack{Below\!Median\!Revenue}   &\Centerstack{Above\!Median\!Revenue}   \\
\hline Observations &      76,773   &     146,249   &      76,773   &     146,249   \\
\(R^{2}\)           &         .55   &         .84   &         .72   &         .47   \\
\end{tabular}

    \flushleft 
    \begin{small}
    \begin{singlespace}
	\textbf{Notes}: This table shows results on lobbying and PAC donations using our panel event study specification in described in Section \ref{sec:panelevent}. We include results separated by firm size measured by revenue. In particular, we sum all revenue across the entire sample for each composite firm, and examine companies above and below the median on this dimension. For additional discussion of this specification, see Section \ref{sec:panelevent}. 
	\end{singlespace}
	\end{small}
\end{table}

\begin{table}[htbp]\centering
\caption{\textbf{Heterogeneity: Close vs Distant Mergers (Panel Event Study)} \label{tab:CloseVsDistant}}

\begin{tabular}{l*{14}{c}}
                    &\multicolumn{1}{c}{(1)}&\multicolumn{1}{c}{(2)}&\multicolumn{1}{c}{(3)}&\multicolumn{1}{c}{(4)}\\
                    &\multicolumn{1}{c}{\Centerstack{Lobby\!Amount}}&\multicolumn{1}{c}{\Centerstack{Lobby\!Amount}}&\multicolumn{1}{c}{\Centerstack{PAC\!Contribs}}&\multicolumn{1}{c}{\Centerstack{PAC\!Contribs}}\\
\hline
\# Component Firms  &     -91,572** &     -91,351** &      -3,027   &      -2,480   \\
                    &    (41,214)   &    (35,909)   &     (2,555)   &     (2,160)   \\
\# Component Firms $\times$ Unique NAICS&       8,204** &       8,360** &         105   &          70   \\
                    &     (3,816)   &     (3,866)   &       (207)   &       (146)   \\
\hline
Additional Controls &               &           Y   &               &           Y   \\
\hline Observations &     223,043   &     223,022   &     223,043   &     223,022   \\
\(R^{2}\)           &         .79   &         .83   &         .32   &         .48   \\
\end{tabular}

    \flushleft 
    \begin{small}
    \begin{singlespace}
	\textbf{Notes}: This table shows results on lobbying and PAC donations using our panel event study specification in described in Section \ref{sec:panelevent}. We include interactions with how many industries are included among the merging firms using NAICS codes. For additional discussion of this specification, see Section \ref{sec:panelevent}. 
	\end{singlespace}
	\end{small}
\end{table}

\begin{table}[htbp]\centering
\caption{\textbf{Exposure Design Results} \label{tab:ExposureDesign}}

\begin{center}
    \flushleft 
    \emph{Panel A: Implementation \#1, $K_{i0}$ = Initial \# of component firms} \\ 
    \begin{tabular}{l*{14}{c}}
                    &\multicolumn{1}{c}{(1)}&\multicolumn{1}{c}{(2)}&\multicolumn{1}{c}{(3)}&\multicolumn{1}{c}{(4)}\\
                    &\multicolumn{1}{c}{\Centerstack{Lobby\!Amount}}&\multicolumn{1}{c}{\Centerstack{Lobby\!Amount}}&\multicolumn{1}{c}{\Centerstack{PAC\!Contribs}}&\multicolumn{1}{c}{\Centerstack{PAC\!Contribs}}\\
\hline
\# Component Firms  &    -106,615** &    -101,684***&      -9,497*  &      -9,456   \\
                    &    (42,297)   &    (37,871)   &     (5,494)   &     (6,364)   \\
\hline
Controls            &               &           Y   &               &           Y   \\
F-Statistic         &       1,361   &       1,213   &       1,361   &       1,213   \\
\hline Observations &     221,994   &     221,994   &     221,994   &     221,994   \\
\end{tabular}


    \vspace{2mm}
    \flushleft 
    \emph{Panel B: Implementation \#2, $K_{i0}$ = NAICS4 Industry Avg \# of component firms} \\ 
    \begin{tabular}{l*{14}{c}}
                    &\multicolumn{1}{c}{(1)}&\multicolumn{1}{c}{(2)}&\multicolumn{1}{c}{(3)}&\multicolumn{1}{c}{(4)}\\
                    &\multicolumn{1}{c}{\Centerstack{Lobby\!Amount}}&\multicolumn{1}{c}{\Centerstack{Lobby\!Amount}}&\multicolumn{1}{c}{\Centerstack{PAC\!Contribs}}&\multicolumn{1}{c}{\Centerstack{PAC\!Contribs}}\\
\hline
\# Component Firms  &    -140,740*  &    -164,789** &     -18,967   &     -18,868   \\
                    &    (83,248)   &    (77,418)   &    (12,266)   &    (14,178)   \\
\hline
Controls            &               &           Y   &               &           Y   \\
F-Statistic         &          91   &          22   &          91   &          22   \\
\hline Observations &     221,994   &     221,994   &     221,994   &     221,994   \\
\end{tabular}

\end{center}

    \flushleft 
    \begin{small}
    \begin{singlespace}
	\textbf{Notes}: This table shows results on lobbying and PAC donations using our exposure specification in described in Section \ref{sec:exposiv}. 
	\end{singlespace}
	\end{small}
\end{table}

\begin{table}[htbp]\centering
\caption{\textbf{Firm-Level Political Risk} \label{tab:PolRisk}}

\flushleft 
    \emph{Panel A: Panel Event Study} \\ 
\begin{tabular}{l*{14}{c}}
                    &\multicolumn{1}{c}{(1)}&\multicolumn{1}{c}{(2)}&\multicolumn{1}{c}{(3)}&\multicolumn{1}{c}{(4)}&\multicolumn{1}{c}{(5)}\\
                    &\multicolumn{1}{c}{\Centerstack{Lobby\!Amount}}&\multicolumn{1}{c}{\Centerstack{PAC\!Contribs}}&\multicolumn{1}{c}{\Centerstack{Political\!Risk}}&\multicolumn{1}{c}{\Centerstack{Econ. Policy\!Political Risk}}&\multicolumn{1}{c}{\Centerstack{Political\!Sentiment}}\\
\hline
\# Component Firms  &     -40,267   &      -1,539   &      -.0043   &      -.0069   &      -.0099   \\
                    &    (27,186)   &     (1,643)   &     (.0082)   &     (.0077)   &       (.01)   \\
\hline
Additional Controls &           Y   &           Y   &           Y   &           Y   &           Y   \\
\hline Observations &      69,789   &      69,789   &      69,789   &      69,789   &      69,789   \\
\(R^{2}\)           &         .88   &         .51   &         .59   &         .58   &          .6   \\
\end{tabular}
 
\\
\vspace{10mm} 
    \emph{Panel B: Exposure Design, Implementation \#1, $K_{i0}$ = Initial \# of component firms} \\ 
\begin{tabular}{l*{14}{c}}
                    &\multicolumn{1}{c}{(1)}&\multicolumn{1}{c}{(2)}&\multicolumn{1}{c}{(3)}&\multicolumn{1}{c}{(4)}&\multicolumn{1}{c}{(5)}\\
                    &\multicolumn{1}{c}{\Centerstack{Lobby\!Amount}}&\multicolumn{1}{c}{\Centerstack{PAC\!Contribs}}&\multicolumn{1}{c}{\Centerstack{Political\!Risk}}&\multicolumn{1}{c}{\Centerstack{Econ. Policy\!Political Risk}}&\multicolumn{1}{c}{\Centerstack{Political\!Sentiment}}\\
\hline
\# Component Firms  &     -69,007*  &     -10,509   &       -.012   &       -.016   &       -.029** \\
                    &    (40,622)   &     (8,386)   &      (.011)   &      (.011)   &      (.014)   \\
\hline
Controls            &           Y   &           Y   &           Y   &           Y   &           Y   \\
F-Statistic         &         276   &         276   &         276   &         276   &         276   \\
\hline Observations &      69,456   &      69,456   &      69,456   &      69,456   &      69,456   \\
\end{tabular}

    \flushleft 
    \begin{small}
    \begin{singlespace}
	\textbf{Notes}: This table examines firm-level political risk. We have firm-level political risk scores for approximately 1/3rd of our sample using the method in \cite{hassan2019firm}. We use these values as outcomes. For additional discussion, see Section \ref{sec:altexplan}.
	
	\end{singlespace}
	\end{small}
\end{table}
\end{bibunit}
\clearpage 

\begin{bibunit}
\appendix
\setcounter{table}{0} \renewcommand{\thetable}{\Alph{section}\arabic{table}}
\renewcommand{\thefigure}{\Alph{section}\arabic{figure}}{\bf \huge Appendix: For Online Publication}
\singlespacing
\section{Theoretical Appendix: Proofs of Propositions}\label{app:proofs}

\begin{proof}[Proof of Proposition \ref{prop:bis0}]
For the first part of the proposition, $R^{\ast}$ is given by the value of $R$ that maximizes joint surplus. For the second part, we have
from Corollary 1 that, in a coalition with only a duopolist, the policy maker
could select $R$ to maximize $(A+aR)^{2}/9-w_{1}R^{2}/2$ with an interior
solution $R_{\{1\}}^{\ast}=\frac{2Ak_{R}/a}{9-4k_{R}}$. This results in%
\begin{equation}
\hat{t}_{1}\geq\frac{18A^{2}k_{R}}{(9-4k_{R})^{2}(9-2k_{R})}.\label{ineqduo}%
\end{equation}
A similar expression applies to $\hat{t}_{2}\geq\frac{18A^{2}k_{R}}%
{(9-4k_{R})^{2}(9-2k_{R})}$.

The grand coalition (firm 1 + firm 2) will instead need to compensate the policy-maker loss at the policy equilibrium%

A quick inspection tells that (\ref{ineqtotduo}) is the key constraint. Intuitively, (\ref{ineqduo}) is not binding because the alternative coalition without firm 1 will still be choosing a relatively high level of the common $R$, and hence the \textquotedblleft punishment\textquotedblright\ that could be inflicted on firm 1 is limited. Instead, jointly both firms need to convince the regulator to enact a policy choice that is against social welfare: the higher the level of the policy, the more the regulator will have
to be compensated with higher transfers. There is a continuum of contribution pairs that satisfy condition (\ref{ineqtotduo}), which can be thought of as different allocations of contributions to a common good. It is natural to select the symmetric one (this is also the solution that minimizes, e.g., the sum of the square of transfers) to finally get%
\begin{equation}
\hat{t}_{1}=\hat{t}_{2}=\frac{4A^{2}k_{R}}{(9-4k_{R})^{2}}.\label{tduo}%
\end{equation}

\end{proof}

\begin{proof}[Proof of Proposition \ref{prop:ais0}]

Policy is found by maximizing surplus. Turning to the contributions, we have that
for firm 1, in a coalition with the other duopolist, the policy maker could
select the $F_{i}$'s to maximize $\pi_{2}-w(F_{1}^{2}/2+F_{2}^{2}/2)$ with an
interior solution $F_{1\{1\}}^{\ast}=\frac{-2Ak_{F}/b}{9-10k_{F}}%
,F_{2\{1\}}^{\ast}=\frac{4Ak_{F}/b}{9-10k_{F}}$. Note that the
\textquotedblleft punishment\textquotedblright\ to firm 1 is quite harsh (in
fact, a negative $F_{1}$). This results in a minimum transfer for firm 1 of%
\begin{equation}
\hat{t}_{1}\geq\frac{18A^{2}k_{F}(5-2k_{F})}{(9-2k_{F})^{2}(9-10k_{F}%
)}.\label{ineqduobis}%
\end{equation}
A similar expression applies to $\hat{t}_{2}\geq\frac{18A^{2}k_{F}(5-2k_{F}%
)}{(9-2k_{F})^{2}(9-10k_{F})}$.

The grand coalition (firm 1 + firm 2) will instead need to compensate the
policy-maker loss at the policy equilibrium%
\[
\hat{t}_{1}+\hat{t}_{2}\geq w(F_{1}^{\ast2}/2+F_{2}^{\ast2}/2)=\frac
{4A^{2}k_{F}}{(9-2k_{F})^{2}}.
\]

In contrast with the case with only the common lobbying component, with private lobbying components (\ref{ineqduobis}) are now the key constraints. The intuition comes from the harsh punishment illustrated above. Then we get%
\begin{equation}
\hat{t}_{1}=\hat{t}_{2}=\frac{18A^{2}k_{F}(5-2k_{F})}{(9-2k_{F})^{2}%
(9-10k_{F})}.\label{tFduo}%
\end{equation}

\end{proof}

\begin{proof}[Proof of Proposition \ref{prop:merger}]
Consider first the case with only $R$ ($b=0).$ That regulation increases with the merger, follows from comparing $R^{\ast}$ from (\ref{Rduo}) with the
corresponding $R^{m}$ in (\ref{Rm}): $R^{\ast}=\frac{4Ak_{R}/a}{9-4k_{R}%
}<R^{m}=\frac{Ak_{R}/a}{2-k_{R}}\Rightarrow\frac{4}{9-4k_{R}}<\frac{1}%
{2-k_{R}}\Rightarrow9-4k_{R}>8-4k_{R}.$

Transfers also increase, as obtained by comparing $\hat{t}_{m}$ given by
(\ref{tm}) with $\hat{t}_{1}+$ $\hat{t}_{2}$ given by (\ref{tduo}): $\hat
{t}_{1}+\hat{t}_{2}=\frac{8A^{2}k_{R}}{(9-4k_{R})^{2}}<\hat{t}_{m}=\frac
{A^{2}k_{R}}{2(2-k_{R})^{2}}\Rightarrow\frac{1}{(9-4k_{R})^{2}}<\frac
{1}{16(2-k_{R})^{2}}\Rightarrow9-4k_{R}>8-4k_{R}.$

Consider next the case with only $F$ ($a=0).$ We have already discussed that one private component (e.g., $F_{2}$) disappears with the merger. The remaining component $F_{1}$ instead goes up, as can be checked by comparing
(\ref{Rm}) with (\ref{Fduo}): $F_{1}^{\ast}=\frac{2Ak_{F}/b}{9-2k_{F}}%
<F_{1}^{m}=\frac{Ak_{F}/b}{2-k_{F}}\Rightarrow\frac{2}{9-2k_{F}}<\frac
{1}{2-k_{F}}\Rightarrow9-2k_{F}>8-2k_{F}.$

Finally, transfers decrease, as obtained by comparing $\hat{t}_{m}$ given by
(\ref{tm}) with $\hat{t}_{1}+$ $\hat{t}_{2}$ given by (\ref{tFduo}): $\hat
{t}_{1}+\hat{t}_{2}=\frac{36A^{2}k_{F}(5-2k_{F})}{(9-2k_{F})^{2}(9-10k_{F}%
)}>\hat{t}_{m}=\frac{A^{2}k_{F}}{2(2-k_{F})^{2}}\Rightarrow\frac{36(5-2k_{F}%
)}{(9-2k_{F})^{2}(9-10k_{F})}>\frac{1}{2(2-k_{F})^{2}}$, which is always
verified in the range required for an interior solution ($1/k_{R}\equiv
w_{2}/b^{2}>9/2$).\smallskip
\end{proof}

\section{Example of a Composite Firm}\label{ComposFirmExample}

Below we show a visual example of a composite firm that starts off as four distinct component firms (A-D) and merges into one over three periods (half years in our sample). Figure \ref{fig:ExampVis} below shows the evolution of this composite firm from period 1 (top) to period 3 (bottom). 

In this example, all component firms' revenue was \$1 for all periods, and there was no organic growth over the three periods. At the end when all four firms are merged, the final firm is worth \$4. This example keeps size/revenue constant for clarity; our actual data include organic growth. In the example, the $\text{MergerIndex}_{it}$ varies across the three periods, which we can measure either as a reduction in the number of independent, as-yet-unmerged firms within the composite (``\# of component firms''), or as an increase in the HHI index as described in Appendix \ref{app:HHI}. 

\begin{figure}[h]
\caption{\textbf{Graphical Representation of Composite Firm ``ABCD''} \label{fig:ExampVis}}
    \center
    \includegraphics[width=12.5cm]{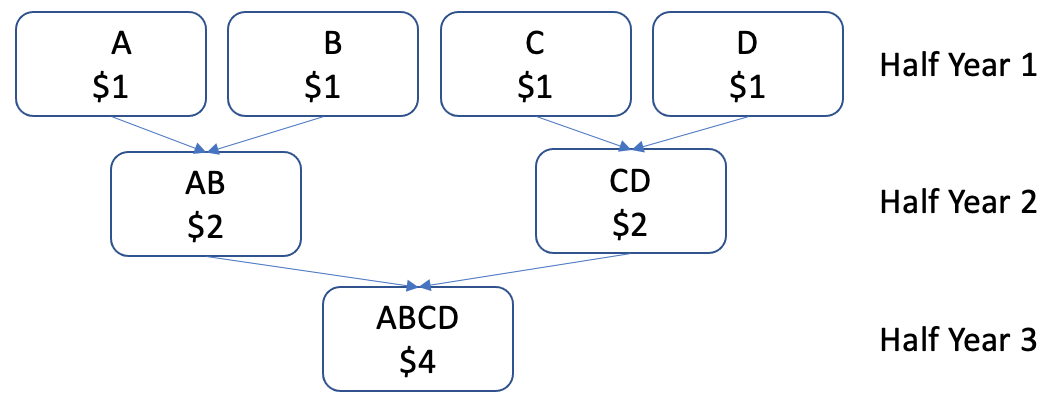}
\end{figure}

\begin{table}[htbp]\centering
\caption{\textbf{Tabular Representation of Figure \ref{fig:ExampVis}, Composite Firm ``ABCD''} \label{fig:ExampTab}}

\resizebox{\textwidth}{!}{
\begin{tabular}{|c|c|c|c|c|c|c|}
     \hline
\multirow{4}{*}{HalfYearID} & \multirow{4}{*}{CompositeFirmID} & \multirow{4}{*}{\Centerstack{Total\! Revenue\!(Size)}} & \multicolumn{3}{c|}{$\text{MergerIndex}_{it}$} \\
\cline{4-6} 
& & & 

\Centerstack{\# of\!Component \!Firms} & \multicolumn{2}{c|}{\Centerstack{HHI\!Index}} \\
    \hline
    1 & ``ABCD'' & \$4 & 4 & 2,500 & =$(1/4)^2\times 4 \times 10K$ \\
    2 & ``ABCD'' & \$4 & 2 & 5,000 & =$(1/2)^2\times 2 \times 10K$ \\
    3 & ``ABCD'' & \$4 & 1 & 10,000 & =$(1/1)^2\times 1 \times 10K$ \\
    \hline 
    \end{tabular}
}
\end{table}

\section{Codifying Multi-Merger Firms}\label{WhyCompFirms}

As described in Section \ref{sec:data}, our composite firm representation is particularly helpful for analyzing multi-merger firms. Mergers are relatively rare. However, among companies that \emph{do} merge with others in our sample, 42\% are involved in multiple mergers or acquisitions. This number rises to 68\% if unlisted companies are included. Multi-merger firms are particularly common among larger companies that may be the source of important political and/or economic influence. Composite firms with more than two components comprise 58\% of all lobbying spend (83\% if  unlisted companies are included). Such firms are often both targets and acquirers in the same sample. 

Multi-merger firms present a data representation challenge. More generally, analysis of networks featuring merging nodes is rare in any network setting. \cite{hernandez2018acquisitions} examines ``node collapse'' through simulations. Our approach of building a ``composite node'' (in our case, a composite firm) for handling this problem may have applications in other empirical settings featuring merging nodes. 

In standard datasets of corporate mergers, target firms disappear after an acquisition. However, the target firm has not disappeared, it has been joined into a larger entity. Some researchers drop the target firm from analysis entirely, and focus only on the outcomes of the acquiring firm (both before and after the merger). This is problematic in settings like our model, where researchers want to study changes in the combined output of comes of both firms (compared to pre-trends before the merger). 

In addition, if one drops a target firm entirely then the target's own prior acquisitions (as an acquirer) would also be dropped. As described above, this would remove a large volume of potentially important activity. One could also keep the targets, and represent them as targets in some acquisitions and acquirers in others. However, the double-appearance of these firms would need to be accounted for in standard error clustering. 

Our composite firm representation addresses these issues. Rather than dropping firms or double-counting them, we create a unit of analysis (the composite firm) that can represent multi-merger firms, single-merging firms and non-merging firms. We can track internal changes to the composition of composite firms over time, and cluster standard errors around these composites.

\section{Merging Data Sources}\label{MergingData}

As described in Section \ref{sec:data}, our dataset brought together four datasets: 1) financial data from Compustat, 2) a dataset about mergers from SDC Platinum, 3) a lobbying dataset from \textit{LobbyView}\footnote{\url{https://www.lobbyview.org/}} \citep{kim2018lobbyview}, and 4) corporate PAC contribution data from the Center for Responsive Politics' \emph{OpenSecrets} project.\footnote{\url{https://www.opensecrets.org/bulk-data/}} 

Below we list additional details about how these datasets were merged together. Our merging mostly used standardized identifiers (GVKEY and CUSIP) with the exception of the text-matching used to incorporate the \emph{OpenSecrets} data. 
\begin{itemize}
    \item[1)] \textit{Compustat} identifies companies both using CUSIP and GVKEY identifiers, thus allowing linkages with other data below using either key. 
    \item[2)] The SDC platinum data identify both target and acquiring companies using CUSIP identifiers. Before integrating this data, we added the composite firm identifiers using the procedure described in Appendix \ref{CompositeGraphProc}. 
    \item[3)] \emph{LobbyView} indexes companies using GVKEY identifiers. We link \emph{LobbyView}'s data with other datasets using the GVKEY/CUSIP crosswalk from Compustat. 
    \item[4)] Unlike \emph{LobbyView}, \emph{OpenSecrets} data does not index companies by a standardized identifier, but by company standardizing company names. We merged this data into the other datasets by using a text matching procedure we validated by manual inspection. 
\end{itemize}

\section{Procedure for Creating the Composite Firm Graph}\label{CompositeGraphProc}

The procedure below takes the SDC Platimum merger dataset described in the main paper (and in Appendix \ref{MergingData} above) and a date. 

We begin by removing all M\&A observations after the specified date. Then we use the SDC data to create a graph that connects all merged firms before that date. Although this graph's edges have a direction (i.e., target $\rightarrow$ acquirer), for our purposes in this section an undirected graph connecting targets and acquirers will suffice. 

We then find the connected components of this graph. A connected component is a maximal connected subgraph. All nodes within the subgraph are reachable from every other node in the subgraph, either directly or through paths. However, all nodes in the component subgraph cannot necessarily reach all nodes in the overall graph. In short, a connected component is an ``island'' of nodes that are interconnected with each other, but not the rest of the graph. 

In our setting, a composite firm is a collection of firms (nodes) that are interconnected to each other by mergers (edges). These connections can either be direct (two firms merging) or through paths (A merging with B, which previously merged with C). The members of these clusters of course typically are not necessarily connected to all other firms (directly or through paths), and thus each cluster of inter-merged firms is an isolated, connected subgraph of the larger merger graph. 

Connected components of a graph can be calculated using efficient, well-known algorithms such as the \cite{hopcroft1973algorithm} algorithm. We used the implementation provided by the igraph  scientific computing package (\citealt{csardi2006igraph}, \url{http://igraph.org}), Version 1.2.6 (published October 6, 2020). 

\section{Anticipation}\label{app:anticip}

As mentioned in Section \ref{sec:panelevent}, a key identification challenge is the possibility of pre-merger increases in lobbying activity. Firms could initiate this form of pre-merger lobbying to influence the merger's review by regulators. Alternatively, firms may anticipate a positive review, and begin coordinating and integrating lobbying activity before the official merger date. To address this, we add terms to Equation (\ref{Spec1}) to capture the change in each composite firm's $\text{MergerIndex}_{it}$ between the current period and one period in the future. We denote these as $\Delta \text{ MergerIndex}_{it}\text{, \emph{t}+1}$. Our additional term measures lobbying one period \textit{ahead} of a merger. Table \ref{tab:Anticip} presents these results. Compared to our results without this term in Table \ref{tab:MainPoliSpend}, we see approximately the same magnitudes.

\begin{table}[htbp]\centering
\caption{\textbf{Merger Anticipation Effects} \label{tab:Anticip}}

\begin{tabular}{l*{14}{c}}
                    &\multicolumn{1}{c}{(1)}&\multicolumn{1}{c}{(2)}&\multicolumn{1}{c}{(3)}&\multicolumn{1}{c}{(4)}\\
                    &\multicolumn{1}{c}{\Centerstack{Lobby\!Amount}}&\multicolumn{1}{c}{\Centerstack{Lobby\!Amount}}&\multicolumn{1}{c}{\Centerstack{PAC\!Contribs}}&\multicolumn{1}{c}{\Centerstack{PAC\!Contribs}}\\
\hline
\# Component Firms  &     -77,995** &     -73,918** &      -4,780*  &      -4,583   \\
                    &    (33,742)   &    (30,123)   &     (2,452)   &     (2,948)   \\
$\Delta$ \# Component Firms, $t+1$&      -9,265   &      -9,859   &        -457   &        -243   \\
                    &    (21,880)   &    (24,727)   &     (1,486)   &     (1,268)   \\
\hline
Additional Controls &               &           Y   &               &           Y   \\
\hline Observations &     210,344   &     210,325   &     210,344   &     210,325   \\
\(R^{2}\)           &         .79   &         .79   &         .32   &         .32   \\
\end{tabular}

    \flushleft 
    \begin{footnotesize}
    \begin{singlespace}
	\textbf{Notes}: This table shows the result of our main specification (Equation \ref{Spec1}) with an anticipation term added as described in Appendix \ref{app:anticip} (immediately above). The additional term, $\Delta \text{ MergerIndex}_{it}\text{, \emph{t}+1}$, measures lobbying one period \textit{ahead} of a merger. 
	\end{singlespace}
	\end{footnotesize}
\end{table}

\section{Diagnostics of the Exposure Design Instruments}\label{app:BartikDiagnostic}

\paragraph{Correlation Tests.} To meet the IV requirements, our instrument must satisfy an exclusion restriction. The requirement is that the merger waves do not affect political spending of the exposed units, except through mergers. Like many identifying assumptions, this cannot be directly tested. \cite{goldsmith2020bartik} suggest an empirical test to validate the instrument: Examine whether initial exposures $K_{i0}$ predict the levels (or differences) of shocks $W_{it}$ from other parts of the economy. 

Table \ref{tab:ivdiag} implements this test. To assesses economic significance, we use regressions with standardized values for both the left- and right-hand side variables. The resulting point estimates are less than one one-hundredth of a standard deviation. Until other controls are added, the $R^{2}$ is less than one ten-thousandth.  This is a correlation of effectively zero in economic significance. Because of our large dataset, we do find statistically significant correlations (our standard errors are even smaller than our point estimates). However, the magnitude of these correlations are effectively zero. 

\paragraph{Compliers \& Instrument Strength.} Compliers to the instrument are composite firms that contain mergers, but whose timing of mergers are sensitive to waves. Many other mergers happen on a timeline unaffected by these waves, or never happen at all; these are not identified by our instrument. In Table \ref{tab:ivcomply}, we assess whether instrument compliance is different by size (measured in revenue). We find that large companies are more likely to be compliers to our instrument; as a result, our IV estimand will capture effects on companies that are larger than the average company in our sample. This property of the instrument also limits our ability to do heterogeneity analysis on the main effects of mergers, because our instrument is weaker for smaller companies. Overall, our instrument has a strong first stage in both implementations, featuring strong $F$ statistics (as measured using the metrics proposed by \citealt{olea2013robust} and \citealt{stock2005testing, kleibergen2006generalized}).

\begin{table}[htbp]\centering
\caption{\textbf{IV Diagnostic: Does Initial Concentration Level Predict Shocks?} \label{tab:ivdiag}}

    \resizebox{\textwidth}{!}{\begin{tabular}{l*{14}{c}}
                    &\multicolumn{1}{c}{(1)}&\multicolumn{1}{c}{(2)}&\multicolumn{1}{c}{(3)}&\multicolumn{1}{c}{(4)}\\
                    &\multicolumn{1}{c}{\Centerstack{Merger Shocks\!(Levels)}}&\multicolumn{1}{c}{\Centerstack{Merger Shocks\!(Changes)}}&\multicolumn{1}{c}{\Centerstack{Merger Shocks\!(Levels)}}&\multicolumn{1}{c}{\Centerstack{Merger Shocks\!(Changes)}}\\
\hline
Component Firms in Period 0&     -.00041   &      .00093** &               &               \\
                    &     (.0003)   &     (.0004)   &               &               \\
Industry Average, Component Firms in Period 0&               &               &       -.004***&       .0046***\\
                    &               &               &    (.00059)   &    (.00064)   \\
Constant            &     5.3e-09   &    -6.4e-06   &     5.3e-09   &    -.000048   \\
                    &    (.00052)   &    (.00074)   &    (.00052)   &    (.00074)   \\
\hline
Controls            &           Y   &           Y   &           Y   &           Y   \\
\hline Observations &     221,994   &     209,390   &     221,994   &     209,390   \\
\(R^{2}\)           &         .99   &         .89   &         .99   &         .89   \\
\end{tabular}
}

    \flushleft 
    \begin{footnotesize}
    \begin{singlespace}
	\textbf{Notes}: All variables have been standardized, and regressions include half-year fixed effects and controls for revenue. 
	\end{singlespace}
	\end{footnotesize}
\end{table}

\begin{table}[htbp]\centering
\caption{\textbf{IV Compliance Heterogeneity: Firm Size in Revenue} \label{tab:ivcomply}}

    \resizebox{\textwidth}{!}{\begin{tabular}{l*{14}{c}}
                    &\multicolumn{1}{c}{(1)}&\multicolumn{1}{c}{(2)}&\multicolumn{1}{c}{(3)}&\multicolumn{1}{c}{(4)}\\
                    &\multicolumn{1}{c}{\Centerstack{\# of Component\!Firms}}&\multicolumn{1}{c}{\Centerstack{\# of Component\!Firms}}&\multicolumn{1}{c}{\Centerstack{\# of Component\!Firms}}&\multicolumn{1}{c}{\Centerstack{\# of Component\!Firms}}\\
\hline
Instrument          &           7***&         7.1***&         1.2***&         1.6***\\
                    &       (.27)   &        (.3)   &       (.18)   &       (.51)   \\
Instrument $\times$ Large Firm&         .35***&         .24*  &         .74***&         .78***\\
                    &        (.1)   &       (.14)   &      (.044)   &      (.056)   \\
\hline
Instrument Version  &         \#1   &         \#1   &         \#2   &         \#2   \\
Controls            &               &           Y   &               &           Y   \\
\hline Observations &     221,498   &     221,498   &     221,498   &     221,498   \\
\(R^{2}\)           &         .86   &         .86   &         .64   &         .67   \\
\end{tabular}
}

    \flushleft 
    \begin{footnotesize}
    \begin{singlespace}
	\textbf{Notes}: All variables have been standardized, and regressions include half-year fixed effects and controls for revenue. 
	\end{singlespace}
	\end{footnotesize}
\end{table}

\section{Details of Investor Call Sample}\label{app:investorcalls}

Our investor call data comes the method developed by \cite{hassan2019firm}. Data from this measure are distributed at \url{https://firmlevelrisk.com}. Political risk measures are available only for the subset of firms that have regular investor calls. The original format of this data indexed in CUSIP identifiers. We merged these into our composite firms format using the following rules. Mergers in our sample fell into three categories: 
\begin{itemize}
    \item[1)] Mergers where all merging firms were in the investor calls. In this case, the composite firm was included in the investor call sample. We measured the overall political risk for the composite firm $i$ at time $t$ as the revenue-weighted average of all the component firms. 
    
    \item[2)] Mergers where some (but not all) of the merging firms held regular investor calls. This occurred when a large firm with regular calls acquired a smaller firm that did not have regular calls. We measured the overall political risk for the composite firm $i$ at time $t$ as the revenue-weighted average of all the component firms that held calls. We included these instances in the investor call sample. 
    
    \item[3)] Finally, mergers where none of the merging firms were in the investor call subsample. We excluded these firms from our investor call sample. 

\end{itemize}
 
Table \ref{tab:PolRiskSample} contains descriptive statistics for firms that are in our investor call sample, compared to ones that are not. 

\begin{table}[htbp]\centering
\caption{\textbf{Descriptive Statistics: Investor Call Sample} \label{tab:PolRiskSample}}

\resizebox{\textwidth}{!}{\begin{tabular}{l*{6}{c}}
                    &\Centerstack{Not in\!Sample}&\Centerstack{In\!Sample}&  Difference   \\
\hline
Years in Sample     &        6.48&       13.59&       -7.11***\\
Avg Revenue (\$10M, per Half Year)&       19.17&      154.38&     -135.21***\\
\hline Lobby Spend (\$1K, per Half Year)&        8.69&      148.90&     -140.21***\\
Lobbied at all (per Half Year)&        0.03&        0.19&       -0.16***\\
Lobbied at all (ever)&        0.07&        0.36&       -0.30***\\
In-House Lobby Spend (\$1K, per Half Year)&        3.85&      104.17&     -100.32***\\
Lobbying Intermediary Spend (\$1K, per Half Year)&        4.85&       44.73&      -39.89***\\
\hline PAC Donations (\$1K, per Half Year)&        0.29&        6.39&       -6.10***\\
PAC Donations$>0$ (per half year)&        0.01&        0.12&       -0.11***\\
PAC Donations$>0$ (Ever)&        0.02&        0.20&       -0.17***\\
\hline Individual Donations (\$1K, per Half Year)&        0.11&        1.61&       -1.50***\\
Individual Donations$>0$ (per Half Year)&        0.02&        0.14&       -0.12***\\
Individual Donations$>0$ (Ever)&        0.13&        0.63&       -0.50***\\
\hline Individual + PAC (\$1K, per Half Year)&        0.40&        7.99&       -7.60***\\
Individual + PAC$>0$ (per Half Year)&        0.03&        0.21&       -0.18***\\
Individual + PAC$>0$ (Ever)&        0.14&        0.64&       -0.50***\\
\hline \# of Component Firms in Compustat&        1.05&        1.65&       -0.60***\\
\hline\end{tabular}
}

    \flushleft 
    \begin{small}
    \begin{singlespace}
	\textbf{Notes}: This table displays simple summary statistics for all composite firms and all periods in our sample for which we have measures of political risk \citep{hassan2019firm}. This is about 1/3rd of our full panel sample. Our panel dataset is described in Section \ref{sec:data}, and  composite firms are defined at the beginning of Section \ref{sec:empoverview}. Section \ref{sec:altexplan} discusses our use of political risk scores. 
	\end{singlespace}
	\end{small}
\end{table}

\begin{table}[htbp]\centering
\caption{\textbf{Firm-Level Political Risk from Earnings Calls (Additional Measures)} \label{tab:PolRiskAll}}

\resizebox{\textwidth}{!}{\begin{tabular}{l*{14}{c}}
                    &\multicolumn{1}{c}{(1)}&\multicolumn{1}{c}{(2)}&\multicolumn{1}{c}{(3)}&\multicolumn{1}{c}{(4)}&\multicolumn{1}{c}{(5)}&\multicolumn{1}{c}{(6)}&\multicolumn{1}{c}{(7)}\\
                    &\multicolumn{1}{c}{\Centerstack{Environment}}&\multicolumn{1}{c}{\Centerstack{Trade}}&\multicolumn{1}{c}{\Centerstack{Institutions}}&\multicolumn{1}{c}{\Centerstack{Health}}&\multicolumn{1}{c}{\Centerstack{Security\!\& Defense}}&\multicolumn{1}{c}{\Centerstack{Taxes}}&\multicolumn{1}{c}{\Centerstack{Technology}}\\
\hline
\# Component Firms  &     -.00059   &       -.013** &      .00092   &       .0016   &       -.011   &      -.0033   &       -.005   \\
                    &     (.0068)   &     (.0065)   &     (.0076)   &     (.0055)   &     (.0071)   &      (.006)   &     (.0094)   \\
\hline
Additional Controls &           Y   &           Y   &           Y   &           Y   &           Y   &           Y   &           Y   \\
\hline Observations &      69,789   &      69,789   &      69,789   &      69,789   &      69,789   &      69,789   &      69,789   \\
\(R^{2}\)           &         .51   &         .46   &         .55   &         .52   &         .54   &         .51   &         .54   \\
\end{tabular}
}
\\
\vspace{2mm} 
\resizebox{\textwidth}{!}{\begin{tabular}{l*{14}{c}}
                    &\multicolumn{1}{c}{(1)}&\multicolumn{1}{c}{(2)}&\multicolumn{1}{c}{(3)}&\multicolumn{1}{c}{(4)}&\multicolumn{1}{c}{(5)}&\multicolumn{1}{c}{(6)}&\multicolumn{1}{c}{(7)}\\
                    &\multicolumn{1}{c}{\Centerstack{Environment}}&\multicolumn{1}{c}{\Centerstack{Trade}}&\multicolumn{1}{c}{\Centerstack{Institutions}}&\multicolumn{1}{c}{\Centerstack{Health}}&\multicolumn{1}{c}{\Centerstack{Security\!\& Defense}}&\multicolumn{1}{c}{\Centerstack{Taxes}}&\multicolumn{1}{c}{\Centerstack{Technology}}\\
\hline
\# Component Firms  &       -.016   &       -.021** &      -.0043   &       .0043   &       -.016   &       -.013   &        -.02   \\
                    &      (.012)   &     (.0094)   &     (.0096)   &      (.013)   &     (.0098)   &     (.0096)   &      (.014)   \\
\hline
Controls            &           Y   &           Y   &           Y   &           Y   &           Y   &           Y   &           Y   \\
F-Statistic         &         276   &         276   &         276   &         276   &         276   &         276   &         276   \\
\hline Observations &      69,456   &      69,456   &      69,456   &      69,456   &      69,456   &      69,456   &      69,456   \\
\end{tabular}
}

    \flushleft 
    \begin{small}
    \begin{singlespace}
	\textbf{Notes}: This table examines firm-level political risk. We have firm-level political risk scores for approximately 1/3rd of our sample using the method in \cite{hassan2019firm}. We use these values as outcomes. For additional discussion, see Section \ref{sec:altexplan}.

	\end{singlespace}
	\end{small}
\end{table}

\FloatBarrier 

\section{Empirical Results using HHI as Merger Index}\label{app:HHI}
In this appendix we employ the Herfindahl-Hirschman Index (HHI) of the composite firm as an alternative measure for $\text{MergerIndex}_{it}$, instead of the simple count of the number of independent firms within each composite firm that we used in the main text. 

The HHI is defined as the sum of the squared relative revenue share of each independent firm within the composite firm, or $\text{HHI}_{it}=10K\sum_{f\in\mathcal{F}_{i,t}}[x_{ft}^{2}]$, where $x_{ft}=r_{ft}/\sum_{f\in\mathcal{F}_{i,t}}r_{ft}$ and $r_{ft}$ is revenue. It is a term that can take values between 0 and 10,000. An example is provided in Table \ref{fig:ExampTab}. When a merger is completed, the number of intermediate parents shrinks, and the revenue share is larger inside the intermediate parent that absorbed one of the firms, resulting in a higher HHI.

Results are shown in the Table below and are qualitatively similar to those in Table 5. Note that an increase in concentration in Table 5 reduces the index of concentration, while now HHI would increase it. 

\paragraph{Implementation Notes.} Recall that $\text{MergerIndex}_{it}$ appeares twice in our exposure design: Once as the variable being instrumented, and again when the instrument itself uses the $\text{MergerIndex}_{it}$ of firms outside the focal firm's industry. In our implementation below, we use $HHI$ as the $\text{MergerIndex}_{it}$ in both cases. 

\begin{table}[htbp]\centering
\caption{\textbf{Results, Panel Event Study} \label{tab:HHIPoliSpend}}

\begin{tabular}{l*{14}{c}}
                    &\multicolumn{1}{c}{(1)}&\multicolumn{1}{c}{(2)}&\multicolumn{1}{c}{(3)}&\multicolumn{1}{c}{(4)}\\
                    &\multicolumn{1}{c}{\Centerstack{Lobby\!Amount}}&\multicolumn{1}{c}{\Centerstack{Lobby\!Amount}}&\multicolumn{1}{c}{\Centerstack{PAC\!Contribs}}&\multicolumn{1}{c}{\Centerstack{PAC\!Contribs}}\\
\hline
HHI                 &           6*  &         7.3*  &         .37   &         .39   \\
                    &       (3.5)   &       (4.1)   &       (.28)   &       (.35)   \\
\hline
Additional Controls &               &           Y   &               &           Y   \\
\hline Observations &     223,043   &     223,022   &     223,043   &     223,022   \\
\(R^{2}\)           &         .79   &         .83   &         .32   &         .47   \\
\end{tabular}

    \flushleft 
    \begin{footnotesize}
    \begin{singlespace}
	\textbf{Notes}: This table shows results on lobbying and PAC donations using our panel event study specification in described in Section \ref{sec:panelevent}. We use $HHI$ (implementation described above) as the merger index. 
	\end{singlespace}
	\end{footnotesize}
\end{table}

\begin{table}[htbp]\centering
\caption{\textbf{Heterogeneity (Firm Size in Revenue): Panel Event Study} \label{tab:HHIPoliSpendSize}}

\begin{tabular}{l*{14}{c}}
                    &\multicolumn{1}{c}{(1)}&\multicolumn{1}{c}{(2)}&\multicolumn{1}{c}{(3)}&\multicolumn{1}{c}{(4)}\\
                    &\multicolumn{1}{c}{\Centerstack{Lobby\!Amount}}&\multicolumn{1}{c}{\Centerstack{Lobby\!Amount}}&\multicolumn{1}{c}{\Centerstack{PAC\!Contribs}}&\multicolumn{1}{c}{\Centerstack{PAC\!Contribs}}\\
\hline
HHI                 &         .65** &          13   &        .014** &         .73   \\
                    &       (.26)   &         (8)   &     (.0069)   &       (.67)   \\
\hline
Additional Controls &           Y   &           Y   &           Y   &           Y   \\
\hline Sample       &\Centerstack{Below\!Median\!Revenue}   &\Centerstack{Above\!Median\!Revenue}   &\Centerstack{Below\!Median\!Revenue}   &\Centerstack{Above\!Median\!Revenue}   \\
\hline Observations &      76,773   &     146,249   &      76,773   &     146,249   \\
\(R^{2}\)           &         .55   &         .84   &         .72   &         .47   \\
\end{tabular}

    \flushleft 
    \begin{footnotesize}
    \begin{singlespace}
	\textbf{Notes}: This table shows results on lobbying and PAC donations using our panel event study specification in described in Section \ref{sec:panelevent}. We use $HHI$ (implementation described above) as the merger index. We include interactions with firm size. To measure size, we use revenue. In particular, we sum all revenue across the entire sample for each composite firm, and examine companies above and below the median on this dimension. For additional discussion of this specification, see Section \ref{sec:panelevent}. 
	\end{singlespace}
	\end{footnotesize}
\end{table}

\begin{table}[htbp]\centering
\caption{\textbf{Close vs Distant Mergers (HHI)} \label{tab:CloseVsDistantHHI}}

\begin{tabular}{l*{14}{c}}
                    &\multicolumn{1}{c}{(1)}&\multicolumn{1}{c}{(2)}&\multicolumn{1}{c}{(3)}&\multicolumn{1}{c}{(4)}\\
                    &\multicolumn{1}{c}{\Centerstack{Lobby\!Amount}}&\multicolumn{1}{c}{\Centerstack{Lobby\!Amount}}&\multicolumn{1}{c}{\Centerstack{PAC\!Contribs}}&\multicolumn{1}{c}{\Centerstack{PAC\!Contribs}}\\
\hline
HHI                 &         5.9   &         6.2   &        -.23   &         -.3   \\
                    &        (10)   &       (9.6)   &       (.47)   &       (.62)   \\
HHI $\times$ Unique NAICS&        -2.6   &        -1.8   &         .28   &         .33   \\
                    &       (9.5)   &       (8.7)   &        (.5)   &       (.62)   \\
\hline
Additional Controls &               &           Y   &               &           Y   \\
\hline Observations &     223,043   &     223,022   &     223,043   &     223,022   \\
\(R^{2}\)           &         .79   &         .83   &         .32   &         .47   \\
\end{tabular}

    \flushleft 
    \begin{footnotesize}
    \begin{singlespace}
	\textbf{Notes}: This table shows results on lobbying and PAC donations using our panel event study specification in described in Section \ref{sec:panelevent}. We include interactions with how many industries are included among the merging firms using NAICS codes. For additional discussion of this specification, see Section \ref{sec:panelevent}. Here we use HHI as the $MergerIndex_{it}$.
	\end{singlespace}
	\end{footnotesize}
\end{table}

\begin{table}[htbp]\centering
\caption{\textbf{Results: Exposure Design} \label{tab:HHIExposureDesign}}
\begin{center}
    
    \begin{tabular}{l*{14}{c}}
                    &\multicolumn{1}{c}{(1)}&\multicolumn{1}{c}{(2)}&\multicolumn{1}{c}{(3)}&\multicolumn{1}{c}{(4)}\\
                    &\multicolumn{1}{c}{\Centerstack{Lobby\!Amount}}&\multicolumn{1}{c}{\Centerstack{Lobby\!Amount}}&\multicolumn{1}{c}{\Centerstack{PAC\!Contribs}}&\multicolumn{1}{c}{\Centerstack{PAC\!Contribs}}\\
\hline
Composite firm HHI  &         234** &         247** &          20   &          21   \\
                    &       (108)   &        (99)   &        (14)   &        (16)   \\
\hline
Controls            &               &           Y   &               &           Y   \\
F-Statistic         &          51   &          50   &          51   &          50   \\
\hline Observations &     221,994   &     221,994   &     221,994   &     221,994   \\
\end{tabular}


\end{center}

    \flushleft 
    \begin{footnotesize}
    \begin{singlespace}
	\textbf{Notes}: This table shows results on lobbying and PAC donations using our exposure specification in described in Section \ref{sec:exposiv}. We use $HHI$ (implementation described above) as the merger index. 
	\end{singlespace}
	\end{footnotesize}
\end{table}

\begin{table}[htbp]\centering
\caption{\textbf{Firm-Level Political Risk (HHI)} \label{tab:PolRiskHHI}}

\flushleft 
    \emph{Panel A: Panel Event Study} \\ 
\begin{tabular}{l*{14}{c}}
                    &\multicolumn{1}{c}{(1)}&\multicolumn{1}{c}{(2)}&\multicolumn{1}{c}{(3)}&\multicolumn{1}{c}{(4)}&\multicolumn{1}{c}{(5)}\\
                    &\multicolumn{1}{c}{\Centerstack{Lobby\!Amount}}&\multicolumn{1}{c}{\Centerstack{PAC\!Contribs}}&\multicolumn{1}{c}{\Centerstack{Political\!Risk}}&\multicolumn{1}{c}{\Centerstack{Econ. Policy\!Political Risk}}&\multicolumn{1}{c}{\Centerstack{Political\!Sentiment}}\\
\hline
HHI                 &          24   &         .49   &     3.4e-06   &     6.6e-06   &    -4.0e-06   \\
                    &        (21)   &       (1.2)   &   (6.6e-06)   &   (6.2e-06)   &   (6.8e-06)   \\
\hline
Additional Controls &           Y   &           Y   &           Y   &           Y   &           Y   \\
\hline Observations &      69,789   &      69,789   &      69,789   &      69,789   &      69,789   \\
\(R^{2}\)           &         .88   &         .51   &         .59   &         .58   &          .6   \\
\end{tabular}
 
\\
\vspace{10mm}
    \emph{Panel B: Exposure Design, Implementation \#1, $K_{i0}$ = Initial \# of component firms} \\ 
\begin{tabular}{l*{14}{c}}
                    &\multicolumn{1}{c}{(1)}&\multicolumn{1}{c}{(2)}&\multicolumn{1}{c}{(3)}&\multicolumn{1}{c}{(4)}&\multicolumn{1}{c}{(5)}\\
                    &\multicolumn{1}{c}{\Centerstack{Lobby\!Amount}}&\multicolumn{1}{c}{\Centerstack{PAC\!Contribs}}&\multicolumn{1}{c}{\Centerstack{Political\!Risk}}&\multicolumn{1}{c}{\Centerstack{Econ. Policy\!Political Risk}}&\multicolumn{1}{c}{\Centerstack{Political\!Sentiment}}\\
\hline
Composite firm HHI  &         124*  &          19   &     .000021   &     .000029   &     .000051** \\
                    &        (68)   &        (15)   &    (.00002)   &    (.00002)   &   (.000026)   \\
\hline
Controls            &           Y   &           Y   &           Y   &           Y   &           Y   \\
F-Statistic         &          51   &          51   &          51   &          51   &          51   \\
\hline Observations &      69,456   &      69,456   &      69,456   &      69,456   &      69,456   \\
\end{tabular}

    \flushleft 
    \begin{small}
    \begin{singlespace}
	\textbf{Notes}: This table examines firm-level political risk with HHI as the $MergerIndex_{it}$. We have firm-level political risk scores for approximately 1/3rd of our sample using the method in \cite{hassan2019firm}. We use these values as outcomes. For additional discussion, see Section \ref{sec:altexplan}.
		\end{singlespace}
	\end{small}
\end{table}

\begin{table}[htbp]\centering
\caption{\textbf{Firm-Level Political Risk from Earnings Calls (Additional Measures, HHI)} \label{tab:PolRiskAllHHI}}

\resizebox{\textwidth}{!}{\begin{tabular}{l*{14}{c}}
                    &\multicolumn{1}{c}{(1)}&\multicolumn{1}{c}{(2)}&\multicolumn{1}{c}{(3)}&\multicolumn{1}{c}{(4)}&\multicolumn{1}{c}{(5)}&\multicolumn{1}{c}{(6)}&\multicolumn{1}{c}{(7)}\\
                    &\multicolumn{1}{c}{\Centerstack{Environment}}&\multicolumn{1}{c}{\Centerstack{Trade}}&\multicolumn{1}{c}{\Centerstack{Institutions}}&\multicolumn{1}{c}{\Centerstack{Health}}&\multicolumn{1}{c}{\Centerstack{Security\!\& Defense}}&\multicolumn{1}{c}{\Centerstack{Taxes}}&\multicolumn{1}{c}{\Centerstack{Technology}}\\
\hline
HHI                 &    -9.3e-07   &      .00001*  &    -1.9e-06   &     6.6e-06   &     3.3e-06   &    -1.8e-06   &     4.0e-06   \\
                    &   (5.2e-06)   &   (5.5e-06)   &   (5.1e-06)   &   (8.4e-06)   &   (5.9e-06)   &   (5.6e-06)   &   (6.1e-06)   \\
\hline
Additional Controls &           Y   &           Y   &           Y   &           Y   &           Y   &           Y   &           Y   \\
\hline Observations &      69,789   &      69,789   &      69,789   &      69,789   &      69,789   &      69,789   &      69,789   \\
\(R^{2}\)           &         .51   &         .46   &         .55   &         .52   &         .54   &         .51   &         .54   \\
\end{tabular}
}
\\
\vspace{2mm}
\resizebox{\textwidth}{!}{\begin{tabular}{l*{14}{c}}
                    &\multicolumn{1}{c}{(1)}&\multicolumn{1}{c}{(2)}&\multicolumn{1}{c}{(3)}&\multicolumn{1}{c}{(4)}&\multicolumn{1}{c}{(5)}&\multicolumn{1}{c}{(6)}&\multicolumn{1}{c}{(7)}\\
                    &\multicolumn{1}{c}{\Centerstack{Environment}}&\multicolumn{1}{c}{\Centerstack{Trade}}&\multicolumn{1}{c}{\Centerstack{Institutions}}&\multicolumn{1}{c}{\Centerstack{Health}}&\multicolumn{1}{c}{\Centerstack{Security\!\& Defense}}&\multicolumn{1}{c}{\Centerstack{Taxes}}&\multicolumn{1}{c}{\Centerstack{Technology}}\\
\hline
Composite firm HHI  &     .000028   &     .000037** &     7.7e-06   &    -7.7e-06   &     .000028   &     .000024   &     .000036   \\
                    &   (.000022)   &   (.000016)   &   (.000017)   &   (.000023)   &   (.000018)   &   (.000017)   &   (.000025)   \\
\hline
Controls            &           Y   &           Y   &           Y   &           Y   &           Y   &           Y   &           Y   \\
F-Statistic         &          51   &          51   &          51   &          51   &          51   &          51   &          51   \\
\hline Observations &      69,456   &      69,456   &      69,456   &      69,456   &      69,456   &      69,456   &      69,456   \\
\end{tabular}
}

    \flushleft 
    \begin{small}
    \begin{singlespace}
	\textbf{Notes}: This table examines additional measures of firm-level political risk using HHI as the $MergerIndex_{it}$. We have firm-level political risk scores for approximately 1/3rd of our sample using the method in \cite{hassan2019firm}. We use these values as outcomes. For additional discussion, see Section \ref{sec:altexplan}.

	\end{singlespace}
	\end{small}
\end{table}

\clearpage \putbib

\end{bibunit}

@techreport{callanderfoartasugaya,
  title={Market Competition and Political Influence: An Integrated Approach},
  author={Callander, Steven and Foarta, Dana and Sugaya, Takuo},
  year={2021},
  institution={CEPR Discussion Paper 16210}
}

@techreport{hartscottrodino,
  title={Hart-Scott-Rodino Annual Report: Fiscal Year 2019},
  author={Simons, Joseph J and Delrahim, Makan},
  year={2020},
  institution={Federal Trade Commission and Department of Justice}
}

@article{baron1995integrated,
  title={Integrated strategy: Market and nonmarket components},
  author={Baron, David P},
  journal={California management review},
  volume={37},
  number={2},
  pages={47--65},
  year={1995},
  publisher={SAGE Publications Sage CA: Los Angeles, CA}
}

@article{feldman2021synergy,
  title={Synergy in mergers and acquisitions: Typology, lifecycles, and value},
  author={Feldman, Emilie R and Hernandez, Exequiel},
  journal={Academy of Management Review},
  number={ja},
  year={2021}
}

@article{canen2022political,
  title={Political Distortions, State Capture, and Economic Development in Africa},
  author={Canen, Nathan and Wantchekon, Leonard},
  journal={Journal of Economic Perspectives},
  volume={36},
  number={1},
  pages={101--24},
  year={2022}
}

@article{jefferson1789thomas,
  title={From Thomas Jefferson to Francis Hopkinson, 13 March 1789},
  author={Jefferson, Thomas},
  journal={National Archives: Founders Online},
  year={1789}
}

@article{huneeus2018effects,
  title={The Effects of Firms' Lobbying on Resource Misallocation},
  author={Huneeus, Federico and Kim, In Song},
  year={2018},
  publisher={MIT Political Science Department Research Paper}
}

@article{dube2020monopsony,
  title={Monopsony in online labor markets},
  author={Dube, Arindrajit and Jacobs, Jeff and Naidu, Suresh and Suri, Siddharth},
  journal={American Economic Review: Insights},
  volume={2},
  number={1},
  pages={33--46},
  year={2020}
}

@article{azar2020labor,
  title={Labor market concentration},
  author={Azar, Jos{\'e} and Marinescu, Ioana and Steinbaum, Marshall},
  journal={Journal of Human Resources},
  pages={1218--9914R1},
  year={2020},
  publisher={University of Wisconsin Press}
}

@article{harford2011institutional,
  title={Institutional cross-holdings and their effect on acquisition decisions},
  author={Harford, Jarrad and Jenter, Dirk and Li, Kai},
  journal={Journal of Financial Economics},
  volume={99},
  number={1},
  pages={27--39},
  year={2011},
  publisher={Elsevier}
}

@article{gaspar2005shareholder,
  title={Shareholder investment horizons and the market for corporate control},
  author={Gaspar, Jos{\'e}-Miguel and Massa, Massimo and Matos, Pedro},
  journal={Journal of financial economics},
  volume={76},
  number={1},
  pages={135--165},
  year={2005},
  publisher={Elsevier}
}

@article{ansolabeheredefiguesnyder,
  title={Why is There so Little Money in U.S. Politics?},
  author={Ansolabehere, Stephen and de Figueiredo, John M. and Snyder, James M.},
  journal={Journal of Economics Perspectives},
  volume={17},
  number={1},
  pages={105--130},
  year={2003}
}

@article{bena2014corporate,
  title={Corporate innovations and mergers and acquisitions},
  author={Bena, Jan and Li, Kai},
  journal={The Journal of Finance},
  volume={69},
  number={5},
  pages={1923--1960},
  year={2014},
  publisher={Wiley Online Library}
}

@book{cageprice,
  title={The Price of Democracy:
 How Money Shapes Politics and What to Do about It},
  author={Cage, Julia},
  year={2020},
  publisher={Harvard University Press}
}

@book{grossmanhelpmanbook,
  title={Special Interest Politics},
  author={Grossman, Gene M and Helpman, Elhanan},
  year={2002},
  publisher={MIT Press}
}

@article{gort1969economic,
  title={An economic disturbance theory of mergers},
  author={Gort, Michael},
  journal={The Quarterly Journal of Economics},
  pages={624--642},
  year={1969},
  publisher={JSTOR}
}

@article{kim2017political,
  title={Political cleavages within industry: Firm-level lobbying for trade liberalization},
  author={Kim, In Song},
  journal={American Political Science Review},
  volume={111},
  number={1},
  pages={1--20},
  year={2017},
  publisher={Cambridge University Press}
}

@article{maksimovic2001market,
  title={The market for corporate assets: Who engages in mergers and asset sales and are there efficiency gains?},
  author={Maksimovic, Vojislav and Phillips, Gordon},
  journal={The journal of finance},
  volume={56},
  number={6},
  pages={2019--2065},
  year={2001},
  publisher={Wiley Online Library}
}

@book{weston1990mergers,
  title={Mergers, restructuring, and corporate control},
  author={Weston, J Fred and Chung, Kwang S COEDITOR},
  number={338.8 W5284m Ej. 1},
  year={1990},
  publisher={Prentice Hall,}
}

@book{nelson1959merger,
  title={Merger movements in American industry, 1895-1956},
  author={Nelson, Ralph Lowell},
  number={66},
  year={1959},
  publisher={Princeton university press Princeton}
}

@article{goel2010envious,
  title={Do envious CEOs cause merger waves?},
  author={Goel, Anand M and Thakor, Anjan V},
  journal={The Review of Financial Studies},
  volume={23},
  number={2},
  pages={487--517},
  year={2010},
  publisher={Society for Financial Studies}
}

@article{mitchell1996impact,
  title={The impact of industry shocks on takeover and restructuring activity},
  author={Mitchell, Mark L and Mulherin, J Harold},
  journal={Journal of financial economics},
  volume={41},
  number={2},
  pages={193--229},
  year={1996},
  publisher={Elsevier}
}

@article{bonaime2018does,
  title={Does policy uncertainty affect mergers and acquisitions?},
  author={Bonaime, Alice and Gulen, Huseyin and Ion, Mihai},
  journal={Journal of Financial Economics},
  volume={129},
  number={3},
  pages={531--558},
  year={2018},
  publisher={Elsevier}
}

@article{toxvaerd2008strategic,
  title={Strategic merger waves: A theory of musical chairs},
  author={Toxvaerd, Flavio},
  journal={Journal of Economic Theory},
  volume={140},
  number={1},
  pages={1--26},
  year={2008},
  publisher={Elsevier}
}

@article{maksimovic2013private,
  title={Private and public merger waves},
  author={Maksimovic, Vojislav and Phillips, Gordon and Yang, Liu},
  journal={The Journal of Finance},
  volume={68},
  number={5},
  pages={2177--2217},
  year={2013},
  publisher={Wiley Online Library}
}

@article{ahern2014importance,
  title={The importance of industry links in merger waves},
  author={Ahern, Kenneth R and Harford, Jarrad},
  journal={The Journal of Finance},
  volume={69},
  number={2},
  pages={527--576},
  year={2014},
  publisher={Wiley Online Library}
}

@article{harford2005drives,
  title={What drives merger waves?},
  author={Harford, Jarrad},
  journal={Journal of financial economics},
  volume={77},
  number={3},
  pages={529--560},
  year={2005},
  publisher={Elsevier}
}

@article{rhodes2004market,
  title={Market valuation and merger waves},
  author={Rhodes-Kropf, Matthew and Viswanathan, Steven},
  journal={The Journal of Finance},
  volume={59},
  number={6},
  pages={2685--2718},
  year={2004},
  publisher={Wiley Online Library}
}

@article{zingales2017towards,
  title={Towards a political theory of the firm},
  author={Zingales, Luigi},
  journal={Journal of Economic Perspectives},
  volume={31},
  number={3},
  pages={113--30},
  year={2017}
}

@article{grullon2019us,
  title={Are US industries becoming more concentrated?},
  author={Grullon, Gustavo and Larkin, Yelena and Michaely, Roni},
  journal={Review of Finance},
  volume={23},
  number={4},
  pages={697--743},
  year={2019},
  publisher={Oxford University Press}
}

@article{berry2019increasing,
  title={Do increasing markups matter? lessons from empirical industrial organization},
  author={Berry, Steven and Gaynor, Martin and Scott Morton, Fiona},
  journal={Journal of Economic Perspectives},
  volume={33},
  number={3},
  pages={44--68},
  year={2019}
}

@article{autor2020fall,
  title={The fall of the labor share and the rise of superstar firms},
  author={Autor, David and Dorn, David and Katz, Lawrence F and Patterson, Christina and Van Reenen, John},
  journal={The Quarterly Journal of Economics},
  volume={135},
  number={2},
  pages={645--709},
  year={2020},
  publisher={Oxford University Press}
}

@article{haleblian2012exploring,
  title={Exploring firm characteristics that differentiate leaders from followers in industry merger waves: A competitive dynamics perspective},
  author={Haleblian, Jerayr and McNamara, Gerry and Kolev, Kalin and Dykes, Bernadine J},
  journal={Strategic Management Journal},
  volume={33},
  number={9},
  pages={1037--1052},
  year={2012},
  publisher={Wiley Online Library}
}

@article{de2020rise,
  title={The rise of market power and the macroeconomic implications},
  author={De Loecker, Jan and Eeckhout, Jan and Unger, Gabriel},
  journal={The Quarterly Journal of Economics},
  volume={135},
  number={2},
  pages={561--644},
  year={2020},
  publisher={Oxford University Press}
}

@book{philippon2019great,
  title={The great reversal: how America gave up on free markets},
  author={Philippon, Thomas},
  year={2019},
  publisher={Harvard University Press}
}

@article{bernheim1986common,
  title={Common agency},
  author={Bernheim, B Douglas and Whinston, Michael D},
  journal={Econometrica: Journal of the Econometric Society},
  pages={923--942},
  year={1986},
  publisher={JSTOR}
}

@book{baumgartner1998basic,
  title={Basic interests},
  author={Baumgartner, Frank R and Leech, Beth L},
  year={1998},
  publisher={Princeton University Press}
}

@article{olson1965logic,
  title={The logic of collective action: public goods and the theory of groups},
  author={Olson, Mancur},
  year={1965},
  publisher={Harvard University Press}
}

@article{hart2004business,
  title={" Business" Is Not An Interest Group: On the Study of Companies in American National Politics},
  author={Hart, David M},
  journal={Annu. Rev. Polit. Sci.},
  volume={7},
  pages={47--69},
  year={2004},
  publisher={Annual Reviews}
}

@article{barber2014lobbying,
  title={Lobbying and the collective action problem: comparative evidence from enterprise surveys},
  author={Barber, Benjamin and Pierskalla, Jan and Weschle, Simon},
  journal={Business and Politics},
  volume={16},
  number={2},
  pages={221--246},
  year={2014},
  publisher={Cambridge University Press}
}

@article{khan2017ideological,
  title={The ideological roots of America's market power problem},
  author={Khan, Lina M},
  journal={Yale LJF},
  volume={127},
  pages={960},
  year={2017},
  publisher={HeinOnline}
}

@book{brandeis1914other,
  title={Other People's Money: And how the Bankers Use it},
  author={Brandeis, L.D.},
  isbn={9781989708866},
  lccn={14006184},
  series={HeinOnline Legal classics library},
  url={https://books.google.com/books?id=uCpMAAAAIAAJ},
  year={1914},
  publisher={F.A. Stokes}
}

@article{hopcroft1973algorithm,
  title={Algorithm 447: efficient algorithms for graph manipulation},
  author={Hopcroft, John and Tarjan, Robert},
  journal={Communications of the ACM},
  volume={16},
  number={6},
  pages={372--378},
  year={1973},
  publisher={ACM New York, NY, USA}
}

@article{f,
  title={Towards a political theory of the firm},
  author={Zingales, Luigi},
  journal={Journal of Economic Perspectives},
  volume={31},
  number={3},
  pages={113--30},
  year={2017}
}

@article{bertrand2020investing,
  title={Investing in influence: Investors, portfolio firms, and political giving},
  author={Bertrand, Marianne and Bombardini, Matilde and Fisman, Raymond and Trebbi, Francesco and Yegen, Eyub},
  year={2020}
}

@Article{csardi2006igraph,
    title = {The igraph software package for complex network research},
    author = {Gabor Csardi and Tamas Nepusz},
    journal = {InterJournal},
    volume = {Complex Systems},
    pages = {1695},
    year = {2006},
    url = {https://igraph.org},
  }

@article{bertrand2014whom,
  title={Is it whom you know or what you know? An empirical assessment of the lobbying process},
  author={Bertrand, Marianne and Bombardini, Matilde and Trebbi, Francesco},
  journal={American Economic Review},
  volume={104},
  number={12},
  pages={3885--3920},
  year={2014}
}

@article{i2012revolving,
  title={Revolving door lobbyists},
  author={Blanes i Vidal, Jordi and Draca, Mirko and Fons-Rosen, Christian},
  journal={The American Economic Review},
  volume={102},
  number={7},
  pages={3731},
  year={2012},
  publisher={American Economic Association}
}

@article{hassan2019firm,
  title={Firm-level political risk: Measurement and effects},
  author={Hassan, Tarek A and Hollander, Stephan and Van Lent, Laurence and Tahoun, Ahmed},
  journal={The Quarterly Journal of Economics},
  volume={134},
  number={4},
  pages={2135--2202},
  year={2019},
  publisher={Oxford University Press}
}

@article{grossman1994protection,
  title={Protection for sale},
  author={Grossman, Gene M and Helpman, Elhanan},
  journal={The American Economic Review},
  pages={833--850},
  year={1994},
  publisher={JSTOR}
}

@article{altonji2005selection,
  title={Selection on observed and unobserved variables: Assessing the effectiveness of Catholic schools},
  author={Altonji, Joseph G and Elder, Todd E and Taber, Christopher R},
  journal={Journal of political economy},
  volume={113},
  number={1},
  pages={151--184},
  year={2005},
  publisher={The University of Chicago Press}
}

@article{oster2019unobservable,
  title={Unobservable selection and coefficient stability: Theory and evidence},
  author={Oster, Emily},
  journal={Journal of Business \& Economic Statistics},
  volume={37},
  number={2},
  pages={187--204},
  year={2019},
  publisher={Taylor \& Francis}
}

@article{greenstone2020credit,
  title={Do credit market shocks affect the real economy? Quasi-experimental evidence from the great recession and" normal" economic times},
  author={Greenstone, Michael and Mas, Alexandre and Nguyen, Hoai-Luu},
  journal={American Economic Journal: Economic Policy},
  volume={12},
  number={1},
  pages={200--225},
  year={2020}
}

@article{card2009immigration,
  title={Immigration and inequality},
  author={Card, David},
  journal={American Economic Review},
  volume={99},
  number={2},
  pages={1--21},
  year={2009}
}

@article{david2013china,
  title={The China syndrome: Local labor market effects of import competition in the United States},
  author={David, H and Dorn, David and Hanson, Gordon H},
  journal={American Economic Review},
  volume={103},
  number={6},
  pages={2121--68},
  year={2013}
}

@article{bartik1991benefits,
  title={Who benefits from state and local economic development policies?},
  author={Bartik, Timothy J},
  year={1991},
  publisher={WE Upjohn Institute for Employment Research Kalamazoo, MI}
}

@techreport{borusyak2020non,
  title={Non-random exposure to exogenous shocks: Theory and applications},
  author={Borusyak, Kirill and Hull, Peter},
  year={2020},
  institution={National Bureau of Economic Research}
}

@article{breuer2021bartik,
  title={Bartik Instruments: An Applied Introduction},
  author={Breuer, Matthias},
  journal={Available at SSRN 3786229},
  year={2021}
}

@article{goldsmith2020bartik,
  title={Bartik instruments: What, when, why, and how},
  author={Goldsmith-Pinkham, Paul and Sorkin, Isaac and Swift, Henry},
  journal={American Economic Review},
  volume={110},
  number={8},
  pages={2586--2624},
  year={2020}
}

@article{athey2022design,
  title={Design-based analysis in difference-in-differences settings with staggered adoption},
  author={Athey, Susan and Imbens, Guido W},
  journal={Journal of Econometrics},
  volume={226},
  number={1},
  pages={62--79},
  year={2022},
  publisher={Elsevier}
}

@article{de2020two,
  title={Two-way fixed effects estimators with heterogeneous treatment effects},
  author={De Chaisemartin, Cl{\'e}ment and d'Haultfoeuille, Xavier},
  journal={American Economic Review},
  volume={110},
  number={9},
  pages={2964--96},
  year={2020}
}

@techreport{freyaldenhoven2021visualization,
  title={Visualization, Identification, and Estimation in the Linear Panel Event-Study Design},
  author={Freyaldenhoven, Simon and Hansen, Christian and P{\'e}rez, Jorge P{\'e}rez and Shapiro, Jesse M},
  year={2021},
  institution={National Bureau of Economic Research}
}

@article{matvos2008cross,
  title={Cross-ownership, returns, and voting in mergers},
  author={Matvos, Gregor and Ostrovsky, Michael},
  journal={Journal of Financial Economics},
  volume={89},
  number={3},
  pages={391--403},
  year={2008},
  publisher={Elsevier}
}

@article{bollaert2015securities,
  title={Securities Data Company and Zephyr, data sources for M\&A research},
  author={Bollaert, Helen and Delanghe, Marieke},
  journal={Journal of Corporate Finance},
  volume={33},
  pages={85--100},
  year={2015},
  publisher={Elsevier}
}

@techreport{blonigen2016evidence,
  title={Evidence for the effects of mergers on market power and efficiency},
  author={Blonigen, Bruce A and Pierce, Justin R},
  year={2016},
  institution={National Bureau of Economic Research}
}

@article{rossi2004cross,
  title={Cross-country determinants of mergers and acquisitions},
  author={Rossi, Stefano and Volpin, Paolo F},
  journal={Journal of Financial Economics},
  volume={74},
  number={2},
  pages={277--304},
  year={2004},
  publisher={Elsevier}
}

@article{goldberg1999protection,
  title={Protection for sale: An empirical investigation},
  author={Goldberg, Pinelopi Koujianou and Maggi, Giovanni},
  journal={American Economic Review},
  volume={89},
  number={5},
  pages={1135--1155},
  year={1999}
}

@article{hillman1982declining,
  title={Declining industries and political-support protectionist motives},
  author={Hillman, Arye L},
  journal={The American Economic Review},
  volume={72},
  number={5},
  pages={1180--1187},
  year={1982},
  publisher={JSTOR}
}

@article{tullock1967welfare,
  title={The welfare costs of tariffs, monopolies, and theft},
  author={Tullock, Gordon},
  journal={Economic inquiry},
  volume={5},
  number={3},
  pages={224--232},
  year={1967},
  publisher={Blackwell Publishing Ltd Oxford, UK}
}

@article{mcchesney1987rent,
  title={Rent extraction and rent creation in the economic theory of regulation},
  author={McChesney, Fred S},
  journal={The Journal of Legal Studies},
  volume={16},
  number={1},
  pages={101--118},
  year={1987},
  publisher={The University of Chicago Law School}
}

@article{stigler1971theory,
  title={The theory of economic regulation},
  author={Stigler, George J},
  journal={The Bell journal of economics and management science},
  pages={3--21},
  year={1971},
  publisher={JSTOR}
}

@article{mehta2020politics,
  title={The politics of M\&A antitrust},
  author={Mehta, Mihir N and Srinivasan, Suraj and Zhao, Wanli},
  journal={Journal of Accounting Research},
  volume={58},
  number={1},
  pages={5--53},
  year={2020},
  publisher={Wiley Online Library}
}

@article{bombardini2020empirical,
  title={Empirical models of lobbying},
  author={Bombardini, Matilde and Trebbi, Francesco},
  journal={Annual Review of Economics},
  volume={12},
  pages={391--413},
  year={2020},
  publisher={Annual Reviews}
}

@article{ellis2018lobbies,
  title={Who lobbies whom? Special interests and hired guns},
  author={Ellis, Christopher J and Groll, Thomas},
  year={2018},
  publisher={CESifo Working Paper}
}

@article{reed2021,
  title={Democracy for Sale: Examining the Effects of Concentration on Lobbying in the United States},
  author={Showalter, Reed},
  year={2021},
  publisher={American Economic Liberties Project}
}

@article{McCarty2021,
  title={Economic Concentration and Political Advocacy, 1999-2017},
  author={McCarty, Nolan and Shahshahani, Sepehr},
  year={2021},
  publisher={Working Paper}
}

@article{fidrmuc2018antitrust,
  title={Antitrust merger review costs and acquirer lobbying},
  author={Fidrmuc, Jana P and Roosenboom, Peter and Zhang, Eden Quxian},
  journal={Journal of Corporate Finance},
  volume={51},
  pages={72--97},
  year={2018},
  publisher={Elsevier}
}

@article{barnes2014evaluating,
  title={Evaluating the SDC mergers and acquisitions database},
  author={Barnes, Beau Grant and L. Harp, Nancy and Oler, Derek},
  journal={Financial Review},
  volume={49},
  number={4},
  pages={793--822},
  year={2014},
  publisher={Wiley Online Library}
}

@techreport{kim2018lobbyview,
  title={Lobbyview: Firm-level lobbying \& congressional bills database},
  author={Kim, In Song},
  year={2018},
  institution={Working Paper available from http://web. mit. edu/insong/www/pdf/lobbyview. pdf}
}

@techreport{blanga2021lobbying,
  title={Lobbying for Globalization},
  author={Blanga-Gubbay, Michael and Conconi, Paola and Parenti, Mathieu},
  year={2021}
}

@incollection{bombfrontier2021,
  author      = "Bombardini, Matilde and Cutinelli-Rendina, Olimpia and Trebbi, Francesco",
  title       = "Lobbying Behind the Frontier",
  editor      = "Ufuk Akcigit and John Van Reenen",
  booktitle   = "The Oxford Handbook of Innovation",
  publisher   = "Harvard University Press",
  address     = "Harvard",
  year        = "2021"
}

@article{doidge2017us,
  title={The US listing gap},
  author={Doidge, Craig and Karolyi, G Andrew and Stulz, Ren{\'e} M},
  journal={Journal of Financial Economics},
  volume={123},
  number={3},
  pages={464--487},
  year={2017},
  publisher={Elsevier}
}

@article{grullon2015disappearance,
  title={The disappearance of public firms and the changing nature of US industries},
  author={Grullon, Gustavo and Larkin, Yelena and Michaely, Roni},
  journal={available at dx. doi. org/10.2139/ssrn},
  volume={2612047},
  year={2015}
}

@article{hernandez2018acquisitions,
  title={Acquisitions, node collapse, and network revolution},
  author={Hernandez, Exequiel and Menon, Anoop},
  journal={Management Science},
  volume={64},
  number={4},
  pages={1652--1671},
  year={2018},
  publisher={INFORMS}
}

@book{wu2018curse,
  title={The Curse of Bigness: Antitrust in the New Gilded Age},
  author={Wu, T.},
  isbn={9780999745465},
  lccn={2018949786},
  series={Columbia global reports},
  url={https://books.google.com/books?id=30sLtAEACAAJ},
  year={2018},
  publisher={Columbia Global Reports}
}

@article{pitofsky1978political,
  title={Political Content of Antitrust},
  author={Pitofsky, Robert},
  journal={U. Pa. L. Rev.},
  volume={127},
  pages={1051},
  year={1978},
  publisher={HeinOnline}
}

@article{kleibergen2006generalized,
  title={Generalized reduced rank tests using the singular value decomposition},
  author={Kleibergen, Frank and Paap, Richard},
  journal={Journal of econometrics},
  volume={133},
  number={1},
  pages={97--126},
  year={2006},
  publisher={Elsevier}
}

@article{stock2005testing,
  title={Testing for Weak Instruments in Linear IV Regression},
  author={Stock, James H and Yogo, Motohiro},
  journal={Identification and Inference for Econometric Models: Essays in Honor of Thomas Rothenberg},
  pages={80},
  year={2005},
  publisher={Cambridge University Press}
}

@article{goodman2021difference,
  title={Difference-in-differences with variation in treatment timing},
  author={Goodman-Bacon, Andrew},
  journal={Journal of Econometrics},
  volume={225},
  number={2},
  pages={254--277},
  year={2021},
  publisher={Elsevier}
}

@article{olea2013robust,
  title={A robust test for weak instruments},
  author={Olea, Jos{\'e} Luis Montiel and Pflueger, Carolin},
  journal={Journal of Business \& Economic Statistics},
  volume={31},
  number={3},
  pages={358--369},
  year={2013},
  publisher={Taylor \& Francis}
}

@article{bombardini2012competition,
  title={Competition and political organization: Together or alone in lobbying for trade policy?},
  author={Bombardini, Matilde and Trebbi, Francesco},
  journal={Journal of International Economics},
  volume={87},
  number={1},
  pages={18--26},
  year={2012},
  publisher={Elsevier}
}

@article{gentzkow2011effect,
  title={The effect of newspaper entry and exit on electoral politics},
  author={Gentzkow, Matthew and Shapiro, Jesse M and Sinkinson, Michael},
  journal={American Economic Review},
  volume={101},
  number={7},
  pages={2980--3018},
  year={2011}
}
\end{document}